\documentclass[lettersize,journal]{IEEEtran}
\usepackage{amsmath,amsfonts}
\usepackage{algorithmic}
\usepackage{algorithm}
\usepackage{makecell}
\usepackage{multirow}
\usepackage{subfigure}
\usepackage{stfloats}
\usepackage{url}
\usepackage{verbatim}
\usepackage{graphicx}
\usepackage{cite}
\usepackage{scalerel}
\usepackage{tikz}
\usetikzlibrary{svg.path}

\usepackage{hyperref} 
\usepackage{pifont}

\newcommand{\cmark}{\ding{51}}

\begin{document}



\title{Enhancing Speaker-Independent Dysarthric Speech Severity Classification with DSSCNet and Cross-Corpus Adaptation}



\author{Arnab Kumar Roy, Hemant Kumar Kathania, and Paban Sapkota
\thanks{Arnab Kumar Roy is with Sikkim Manipal Institute of Technology (SMIT), Sikkim, India - 737136 (e-mail: arnab\_202000152@smit.smu.edu.in).}
\thanks{Hemant Kumar Kathania, and Paban Sapkota are with National Institute of Technology Sikkim, India - 737139 (e-mail: hemant.ece@nitsikkim.ac.in, phec230006@nitsikkim.ac.in).}
}



\maketitle

\begin{abstract}
Dysarthric speech severity classification is crucial for objective clinical assessment and progress monitoring in individuals with motor speech disorders. Although prior methods have addressed this task, achieving robust generalization in speaker-independent (SID) scenarios remains challenging. This work introduces DSSCNet, a novel deep neural architecture that combines Convolutional, Squeeze-Excitation (SE), and Residual network, helping it extract discriminative representations of dysarthric speech from mel spectrograms. The addition of SE block selectively focuses on the important features of the dysarthric speech, thereby minimizing loss and enhancing overall model performance. We also propose a cross-corpus fine-tuning framework for severity classification, adapted from detection-based transfer learning approaches. 
DSSCNet is evaluated on two benchmark dysarthric speech corpora: TORGO and UA-Speech under speaker-independent evaluation protocols: One-Speaker-Per-Severity (OSPS) and Leave-One-Speaker-Out (LOSO) protocols. DSSCNet achieves accuracies of 56.84\% and 62.62\%  under OSPS and 63.47\% and 64.18\% under LOSO setting on TORGO and UA-Speech respectively outperforming existing state-of-the-art methods. Upon fine-tuning, the performance improves substantially, with DSSCNet achieving up to 75.80\% accuracy on TORGO and 68.25\% on UA-Speech in OSPS, and up to 77.76\% and 79.44\%, respectively, in LOSO. These results demonstrate the effectiveness and generalizability of DSSCNet for fine-grained severity classification across diverse dysarthric speech datasets.
\end{abstract}


\section{Introduction}
\IEEEPARstart{D}{ysarthria} is a motor speech disorder caused by impaired muscle control, often affecting articulation, phonation, prosody, and overall speech clarity. \cite{Dysarthr64:online}. It is commonly associated with neurological conditions such as stroke, cerebral palsy, Parkinson’s disease, and amyotrophic lateral sclerosis (ALS) \cite{tomik2010dysarthria}. The severity of dysarthria can differ significantly from one individual to another ranging from mild cases, where speech remains largely intelligible, to severe forms that render speech nearly incomprehensible. This wide spectrum presents serious communication challenges and can contribute to social withdrawal and a reduced quality of life for those affected. Accurate assessment of dysarthria severity is critical for both clinical practice and technological applications. In clinical settings, it enables speech-language pathologists (SLPs) to design individualized therapy plans, monitor patient progress, and optimize rehabilitation strategies \cite{stipancic2021you, jayaraman2023dysarthria}. From a technological standpoint, severity classification helps improve the adaptability of Automatic Speech Recognition (ASR) systems to dysarthric speech \cite{yeo2023automatic}, making these systems more accessible and user-friendly. It also supports the development of real-time assistive tools such as speech synthesizers, communication aids, and voice interfaces which are crucial in making technology more inclusive for people with speech impairments \cite{sajiha2024automatic, mustafa2014severity}. Despite its importance, dysarthric speech severity classification remains a complex challenge. High inter-speaker variability, class imbalance, and the scarcity of large, labeled datasets continue to limit the effectiveness of existing approaches. As a result, there is a growing need for more robust and generalizable models that can better capture the nuances of dysarthric speech and improve classification accuracy across diverse user demographics.

Recent advancements in dysarthric speech severity classification have been largely fueled by progress in deep learning and speech processing methodologies. Earlier approaches primarily relied on handcrafted acoustic features, such as Mel-Frequency Cepstral Coefficients (MFCCs), Linear Predictive Coding (LPC), and prosodic attributes. These features were used to train classical machine learning models including Support Vector Machines (SVMs) \cite{10157285}, Hidden Markov Models (HMMs) \cite{hmm}, Random Forests (RF) \cite{10157285}, and Gaussian Mixture Models (GMMs) \cite{10.3233/JIFS-189139}. Although these models achieved reasonable performance, they were heavily dependent on manual feature engineering and struggled with challenges such as speaker variability, noise sensitivity, and poor generalization across different severity levels \cite{kent2000dysarthrias, wilson2000acoustic, morris1989vot, fosler2003tutorial}.


To overcome these limitations, recent studies have increasingly turned to deep learning-based models for dysarthric speech severity classification. Convolutional Neural Networks (CNNs) have proven effective in capturing hierarchical spectral representations, as demonstrated in \cite{shih2022dysarthria}, where a CNN-GRU model was used to learn patterns specific to dysarthric speech. Other techniques such as multi-head attention and multi-task learning have also been explored to model severity-related dependencies more effectively \cite{joshy2023dysarthria}. Temporal modeling has benefited from Recurrent Neural Networks (RNNs) and Long Short-Term Memory (LSTM) networks, which improve classification by leveraging sequential context \cite{bhat2020automatic, al2024detection, joshy2021automated}. More recently, self-supervised learning (SSL) approaches such as wav2vec 2.0 \cite{baevski2020wav2vec} and HuBERT \cite{hsu2021hubert} have shown promising results by pretraining models on large-scale unlabeled speech corpora, thereby enhancing feature representation and enabling more effective transfer to downstream classification tasks. Transformer-based architectures have also been explored for dysarthric speech, particularly in ASR tasks. For instance, \cite{shahamiri2023dysarthric} demonstrated that transformer and attention-based models can effectively recognize severely impaired speech using depth optimization and transfer learning. These results suggest the potential of such architectures for severity classification as well.

Despite these advancements, several limitations persist. Generalization across speakers remains a significant challenge, as many models perform well on seen speakers but struggle with unseen individuals, particularly in speaker-independent (SID) settings. The scarcity of dysarthric speech datasets further amplifies this issue. Moreover, class imbalance remains a critical concern, with severe dysarthria cases being underrepresented, leading to biased predictions and reduced model reliability. To mitigate these issues, researchers have investigated domain adaptation \cite{woszczyk20_interspeech}, multi-task learning \cite{xiong2024improving}, and adversarial training \cite{woszczyk20_interspeech}. However, ensuring robust speaker-independent models that generalize well across diverse speech characteristics remains an open research problem, necessitating further advancements in transfer learning, cross-corpus learning, and speaker-adaptive training strategies.

In this paper, we propose DSSCNet (Dysarthric Speech Severity Classification Network), a deep learning model designed to enhance dysarthric speech severity classification. Unlike traditional approaches that rely on dataset-specific training, DSSCNet leverages pretraining on one dysarthric speech corpus and fine-tune on another. This approach enables the model to extract more robust and transferable speech representations, mitigating speaker dependence and enhancing classification accuracy.

Our main contributions are as follows:

\begin{itemize}
    \item Developed a novel deep learning model: DSSCNet comprising of convolutional, squeeze-excitation (SE), and residual network for minimizing misclassification and enhancing performance.
    \item Comprehensive evaluation of our proposed DSSCNet across SID settings: both one-speaker-per-severity (OSPS) and leave-one-speaker-out (LOSO), ensuring a rigorous assessment of model robustness.
    \item DSSCNet achieved state-of-the-art performance and further improved classification accuracy by leveraging transfer learning, including pretraining and fine-tuning strategies, to enhance generalization across datasets.
    \item Benchmarking on TORGO and UA-Speech datasets, demonstrating state-of-the-art performance compared to existing methods, with significant improvements in misclassification reduction.
\end{itemize}

\section{Methodology}
\label{sec:methodology}

\subsection{Proposed Classification Network: DSSCNet}
\label{subsec:dsscnet}
The Dysarthric Speech Severity Classification Network (DSSCNet) comprises three main components: a convolutional backbone for initial feature extraction, a Squeeze-and-Excitation (SE) block for adaptive channel-wise recalibration, and a Residual Network to capture complex hierarchical representations. This architectural composition enhances the model’s ability to minimize training loss and effectively learn discriminative features for accurate severity classification. The overall architecture of DSSCNet is depicted in Fig. \ref{fig:model-arch}.

\subsubsection{Simple Feature Extraction}
\label{subsubsec:cnn}

Convolutional Neural Networks (CNNs) are widely used for low-level feature extraction in speech and audio processing tasks due to their ability to capture localized spectral and temporal patterns. In DSSCNet, a CNN module is employed to enhance learning stability and training efficiency. It consists of three convolutional layers with filter sizes of $64$, $128$, and $256$, enabling progressive abstraction of feature representations. Each convolutional layer is followed by batch normalization to stabilize the learning process and a ReLU activation to introduce non-linearity. Subsequently, max-pooling is applied to reduce spatial dimensions, thereby decreasing computational overhead and introducing translational invariance. Together, these layers form the foundation of the feature extraction pipeline in DSSCNet.

The feature extraction process can be expressed using the CNN module represented as CNet as follows:
\begin{equation}
    X_{\text{FE}} = \text{CNet}(X)
\end{equation}
where $X$ is the mel spectrogram of a speech sample, and $X_{\text{FE}}$ is the output feature map obtained from CNet. This feature map serves as input to the subsequent Squeeze and Excitation (SE) network, which further refines the representations by dynamically reweighting feature channels. This ensures that the most discriminative speech features are emphasized.


\subsubsection{Loss Reduction using Squeeze and Excitation Network}
\label{subsubsec:loss-reduction}

\begin{figure*}[!ht]
    \centering
    \includegraphics[width=\linewidth]{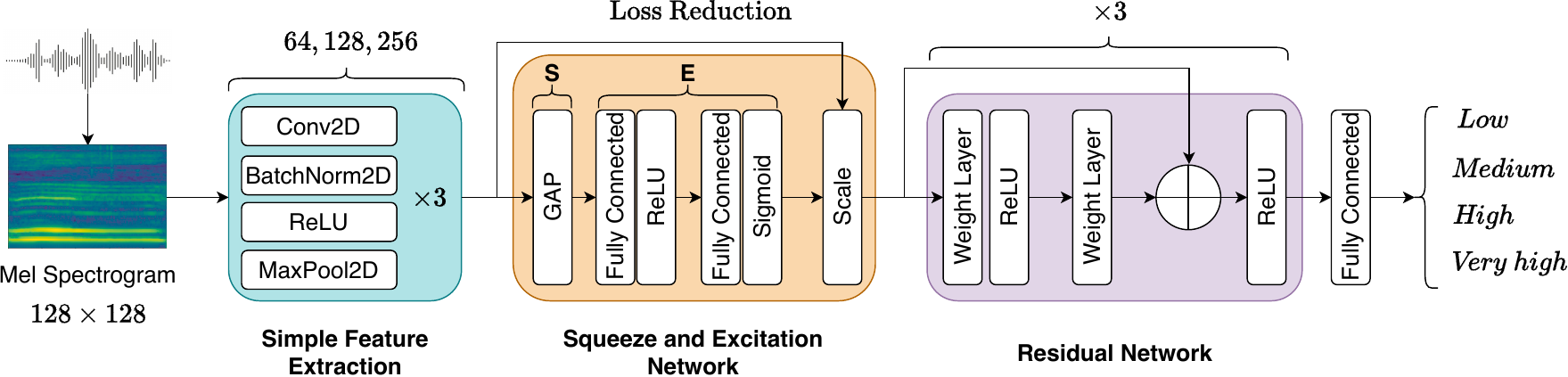}
    \caption{Overview of DSSCNet for dysarthria speech severity classification. S and E refers to squeeze and excitation operations.}
    \label{fig:model-arch}
\end{figure*}



The Squeeze-and-Excitation (SE) network \cite{hu2018squeeze} is a lightweight architectural module designed to improve a neural network’s sensitivity to channel-wise feature importance. Originally proposed for image classification tasks, SE blocks have since been successfully applied to a range of speech-related applications, including emotion recognition \cite{10.1007/978-3-030-98358-1_42, li2024ms}, voice-based disease detection \cite{ai6040068} and even dysarthric speech severity classification \cite{joshy2023biomedical}, due to their ability to emphasize task-relevant spectral characteristics. 

Following these works, we incorporate the SE mechanism into DSSCNet to enhance feature representation and improve channel discriminability. The SE block is positioned after the initial feature extraction module and operates by dynamically re-calibrating channel-wise feature responses, enabling the model to focus on the most informative aspects of the input speech signal while suppressing irrelevant or redundant features.

The SE Network comprises two primary operations: squeeze and excitation. The squeeze operation utilizes Global Average Pooling (GAP) \cite{lin2013network} to aggregate spatial data across each feature map, thereby generating a global feature descriptor that encapsulates channel-wise activations.

The squeeze operation with $X_{\text{FE}}$ as an input can be expressed as:
\begin{equation}
    X_{\text{SQ}} = \frac{1}{H \times W} \sum_{i=1}^{H} \sum_{j=1}^{W} X_{\text{FE}}(i,j)
\end{equation}

Here, $H$ and $W$ are the height and width of the feature map, respectively, and $X_{\text{FE}}(i, j)$ denotes the activation at position $(i, j)$.

In the excitation operation, the global descriptor, denoted as $X_{\text{SQ}}$, undergoes a transformation through two fully connected (FC) layers. The initial layer employs a dimensionality reduction technique, followed by a dimensionality expansion layer, with a rectified linear unit (ReLU) activation function interposed. The subsequent FC layer is followed by a sigmoid activation function to generate per-channel attention weights, represented as:
\begin{equation}
s = \sigma(W_2 \delta(W_1 X_{\text{SQ}}))
\end{equation}

where $W_1 \text{ and } W_2$ are the weight matrices from the FC layers, $\delta(\cdot)$ denotes the ReLU activation function, and $\sigma(\cdot)$ represents the sigmoid activation. The output $s$ contains channel-wise scaling factors, which are used to rescale the original feature maps, reinforcing important speech representations while diminishing irrelevant components:
\begin{equation}
X_{\text{SE}} = s \cdot X_{\text{FE}}
\end{equation}

where $X_{\text{SE}}$ is the reweighted feature map. This adaptive weighting mechanism enhances the model’s sensitivity to dysarthric speech cues by prioritizing features that contribute significantly to severity classification. 

By integrating an SE module, DSSCNet improves its ability to capture subtle variations in dysarthric speech, leading to more accurate severity classification. 
The combination of the CNN-based feature extractor with the squeeze-and-excitation (SE) mechanism forms our baseline model, referred to as CNN + SE similar to \cite{joshy2023biomedical}, which serves as a comparative reference against the proposed DSSCNet architecture throughout our experiments. 
The refined feature maps from the SE network are subsequently passed into the Residual Network Module, where hierarchical speech patterns are further modeled to enhance robustness and generalization.


\subsubsection{Residual Network}
\label{subsubsec:residual}
DSSCNet integrates a Residual Network (ResNet) module \cite{he2016deep} to refine the feature representations obtained from earlier stages of the architecture. Residual learning has proven particularly effective in deep networks, as it preserves essential input information, facilitates gradient propagation, and alleviates issues related to vanishing gradients during training. By incorporating skip connections, DSSCNet is able to model both low-level and high-level dysarthric speech characteristics, which strengthens its ability to discriminate between severity levels in a speaker-independent context.

The residual module comprises a sequence of weight layers interleaved with ReLU activation and skip connections, facilitating direct information transmission. Mathematically, the residual learning operation can be formulated as:

\begin{equation}
X_{\text{Res}} = f(X_{\text{SE}}) + X_{\text{SE}}
\end{equation}

where $X_{\text{SE}}$ is the input feature map from the Squeeze and Excitation (SE) network, $f(X_{\text{SE}})$ represents the transformed feature output of the residual block, and $X_{\text{Res}}$ is the final output. The addition of the skip connection ensures that the original information is retained while allowing the network to learn additional transformations through weight layers. Each residual block in DSSCNet consists of:

\begin{itemize}
    \item \textbf{Weight layers}: Fully Connected (FC) layer that learn additional transformations while preserving essential speech features.
    \item \textbf{ReLU Activation}: Introduces non-linearity, enabling the model to capture complex dysarthric speech variations.
    \item \textbf{Skip Connections}: Enable feature reuse, reducing degradation in deeper layers and improving generalization across speakers and severity levels.
\end{itemize}

The residual block is repeated three times, forming a hierarchical structure that improves representational depth without incurring excessive computational cost. This design is based on empirical validation, where three blocks provided a balanced trade-off between model performance and efficiency, while shallower or deeper variants led to marginal or inconsistent gains. The final output of the residual module is forwarded to a classification layer, which predicts one of four severity levels: Low, Medium, High, or Very High.

\subsection{Framework for Cross-Corpus Fine-Tuning using proposed DSSCNet}
\label{subsec:framework}

\begin{figure}[!ht]
    \centering
    \includegraphics[width=\linewidth]{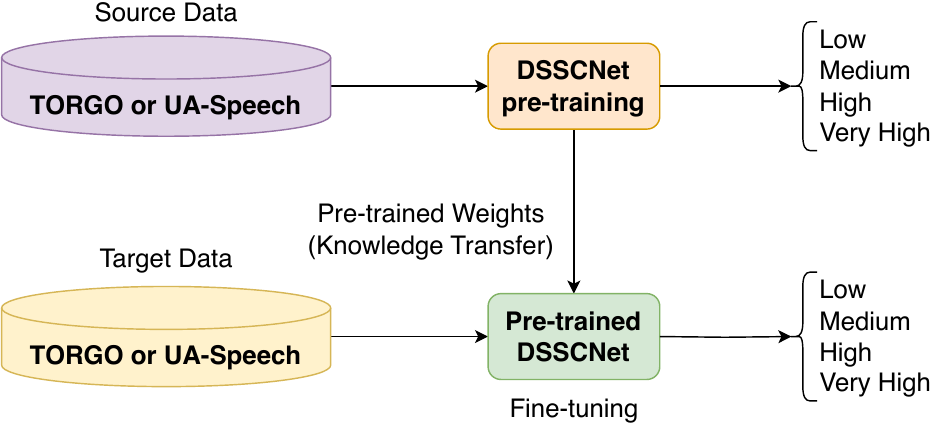}
    \caption{Block diagram of the DSSCNet fine-tuning framework for dysarthric speech severity classification.}
    \label{fig:framework}
\end{figure}

Building upon the foundational concepts of cross-corpus transfer learning explored in prior studies on dysarthria detection \cite{wang2021unsupervised, wav2vecfinetune}, we propose a novel adaptation of this paradigm for the task of dysarthric speech severity classification. The framework, illustrated in Fig. \ref{fig:framework}, leverages the pre-training and fine-tuning strategy to improve generalization and robustness of the proposed DSSCNet model across diverse speech corpora.

In the pre-training phase, DSSCNet is trained on a source corpus to acquire dysarthric speech features that are generalizable. Under speaker-generic constraints, the model can capture the typical patterns associated with severity from the source data. During the fine-tuning process, the pre-trained weights are initially transferred into the target dataset, and then the model is further refined against the dataset’s own features. This approach allows the model to leverage prior knowledge and enhance its generalization capabilities to new speakers and datasets. The final layer of the fine-tuned model is employed to classify the observed dysarthric speech into one of the four severity levels: Low, Medium, High, and Very High.



\section{Dataset and Experimental Setup}
\label{sec:setup}

\subsection{Dataset Description}
\label{subsec:dataset}
This paper employs the TORGO \cite{rudzicz2012torgo} and UA-Speech \cite{kim2008dysarthric} datasets, two publicly available dysarthric speech corpora comprising recordings of individuals exhibiting varying degrees of speech impairment. The speaker-wise severity levels and utterance distribution utilized in this study are presented in Table \ref{tab:dataset}.

\renewcommand{\arraystretch}{1.1}
\begin{table*}[!ht]
    \centering
    \caption{Speaker-wise utterance count and severity level description for both TORGO and UA-Speech dataset.}
    \scalebox{0.94}{
    \begin{tabular}{c|c|ccccc|ccc|ccc|cccc}
        \hline 
    
        \hline
        \multicolumn{2}{c|}{\textbf{Severity}} & \multicolumn{5}{c|}{Low} & \multicolumn{3}{c|}{Medium} & \multicolumn{3}{c|}{High} & \multicolumn{4}{c}{Very High} \\
        \hline

        \multirow{2}{*}{\textbf{TORGO}} & \textbf{Speaker} & F03 & F04 & M03 & - & - & F01 & M05 & - & M01 & M02 & M04 & - & - & - & - \\
        & \textbf{No. of Utterances} & 1075 & 667 & 800 & - & - & 228 & 573 & - & 739 & 766 & 652 & - & - & - & -\\
        \hline

        \multirow{2}{*}{\textbf{UA-Speech}} & \textbf{Speaker} & F05 & M08 & M09 & M10 & M14 & F04 & M05 & M11 & F02 & M07 & M16 & F03 & M04 & M01 & M12 \\
        & \textbf{No. of Utterances} & 5355 & 5355 & 5354 & 5355 & 5355 & 5251 & 5354 & 4590 & 5354 & 5354 & 4590 & 5182 & 3825 & 2805 & 4590 \\
        \hline

        \hline
    \end{tabular}}
    \label{tab:dataset}
\end{table*}

The \textbf{TORGO} \cite{rudzicz2012torgo} dataset comprises recordings from eight dysarthric speakers (three female and five male) exhibiting varying degrees of severity, accompanied by seven age-matched healthy control speakers. Duration of dysarthric speech is $\sim$ 6.2 hours. The dataset encompasses a diverse range of speech samples, including isolated words, sentences, and sustained vowels, captured utilizing both head-mounted and directional microphones at a sampling rate of 16 kHz.

The \textbf{UA-Speech} \cite{kim2008dysarthric} dataset comprises recordings from 15 dysarthric speakers (4 female, 11 male), providing a larger and more diverse corpus for dysarthric speech assessment. Total dysarthric speech duration amounts to $\sim$ 66 hours. We remove leading and trailing silences from each speech sample as part of preprocessing, resulting in a total usable duration of 35.88 hours. The dataset contains a total of 765 unique words, including 300 distinct isolated words, 455 phonetically balanced words, and digit sequences. Each speaker produced multiple repetitions of these words, which were recorded simultaneously using an 8-channel microphone array. Speech samples include isolated words and short phrases, recorded at a 16 kHz sampling rate.

\subsection{Dataset Distribution Across Experimental Settings}
\label{subsec:data-splits}
Assessment of DSSCNet’s ability to generalize across unseen speakers is carried out using SID evaluation protocols, where the test speakers are strictly disjoint from the training set. This evaluation is structured into two configurations:
a) One-Speaker-per-Severity (OSPS), and
b) Leave-One-Speaker-Out (LOSO).

\subsubsection{One-Speaker-per-Severity}
\label{subsubsec:osps-setting}
The OSPS configuration is a SID evaluation protocol designed to assess a model’s ability to generalize to unseen speakers, while ensuring that both the train and test set contains representatives from all severity levels. Specifically, each severity level in the test set is represented by a distinct speaker, and no speaker from the test set appears in the training set, thereby ensuring strict speaker independence.

For OSPS configuration in the TORGO dataset, the dataset was divided into 18 unique speaker combinations using the configurations in \cite{8884185}. 5 dysarthric speakers were allocated for training, while 3 were allocated for testing for each split. This method maintains a balanced severity distribution while preventing speaker overlap between the training and testing sets. The control speakers were excluded from the severity classification task.

For the UA-Speech dataset, out of the total 15 speakers, 12 were selected for the OSPS configuration. This was done to ensure an equitable distribution of speakers across various severity levels. This dataset was divided into 81 unique speaker combinations similar to \cite{8884185}, with 8 dysarthric speakers for training and 4 for testing in each split. This setup enables the model to learn generalized dysarthric speech patterns while minimizing the impact of speaker variability.

\subsubsection{Leave-One-Speaker-Out}
\label{subsubsec:loso-setting}
For a more nuanced generalization assessment, we employ the LOSO evaluation method. In this approach, the model is trained on all but one speaker in the dataset and subsequently tested on the excluded speaker. This procedure is repeated such that each speaker in the dataset serves as the test subject once. Accordingly, this results in 8 distinct speaker-based evaluations for the TORGO dataset and 12 for the UA-Speech dataset, corresponding to the number of dysarthric speakers in each corpus. The final performance is computed as the average across all speaker-specific evaluations, providing a robust estimation of the model’s ability to generalize to entirely unseen speakers, a scenario reflective of real-world deployment conditions.

\subsection{Data Preprocessing and Feature Extraction}
\label{subsec:data-processing}
A consistent and structured preprocessing pipeline is applied to all speech recordings to maintain uniformity in feature extraction and optimize model training. In contrast to previous approaches that averaged feature vectors across frames within each utterance to obtain a fixed-length representation \cite{xiong2024improving}, the current study does not impose explicit truncation or zero-padding, enabling the model to learn from the natural variability in utterance lengths. The speech signal is then transformed into a Mel spectrogram using the Short-Time Fourier Transform (STFT) with an FFT size proportional to a 16 ms window and a 4 ms hop length, ensuring fine-grained temporal resolution. 

Log transformation is applied to the extracted Mel spectrograms to stabilize feature distributions and reduce dynamic range, while preserving key spectral distinctions relevant to dysarthric speech. Each spectrogram is then resized to a fixed $128 \times 128$ resolution using bilinear interpolation, maintaining uniform input dimensions across the dataset. The choice of $128 \times 128$ strikes a balance between temporal resolution and spectral detail, providing sufficient granularity to capture dysarthria-specific articulatory and prosodic patterns, while remaining computationally efficient for training deep convolutional architectures. To match the DSSCNet input format, the single-channel spectrogram is expanded to three channels by replicating the feature maps across three dimensions. This preprocessing strategy ensures that input features are standardized while preserving critical dysarthric speech characteristics, enabling the model to effectively learn severity-related patterns and generalize across speakers with varying speech impairments.

The DSSCNet model is trained using a batch size of 16, a learning rate of $1 \times 10^{-3}$, optimizer of Adam \cite{kingma2014adam}, and for 10 epochs. To optimize the performance of severity classification, the model is trained with the CrossEntropy loss function \cite{zhang2018generalized}, with class-specific weighting applied to address class imbalance. The weights are computed based on the number of samples in each severity class, ensuring that underrepresented classes contribute proportionally to the loss and improving the model’s ability to learn from all severity levels. The CrossEntropy loss measures the difference between the predicted class probabilities and the ground-truth labels, ensuring that the model effectively learns discriminative severity-related features. The loss function is formulated as:

\begin{equation}
\mathcal{L}_{\text{CE}} = - \sum_{i=1}^{C} y_i \log(\hat{y}_i)
\end{equation}

where $C$ represents the number of severity classes, $y_i$ is the ground-truth label, and $\hat{y}_i$ is the predicted probability for class $i$. The Adam optimizer is used to update model weights, leveraging adaptive learning rates to improve convergence speed and stability. This training strategy enables DSSCNet to learn robust dysarthric speech representations, ultimately enhancing classification performance.

\subsection{Transfer Learning with proposed DSSCNet}
\label{subsec:pretraining}
To improve performance in SID configurations such as OSPS and LOSO, we adopt a cross-corpus transfer-learning strategy as discussed in Section \ref{subsec:framework}. DSSCNet is first pre-trained on the complete training dataset of one corpus either TORGO or UA-Speech to learn generalized representations of dysarthric speech across diverse severity levels and speaker characteristics. The pre-train model is then fine-tuned and evaluated on the SID sets of the other corpus, enabling us to assess the model's ability to generalize to unseen speakers under different recording and linguistic conditions.

To ensure clarity and reproducibility, the following two cross-corpus configurations are evaluated:

\subsubsection{UA-Speech $\rightarrow$ TORGO}
\label{subsubsec:pretrained-uaspeech}
DSSCNet is pre-trained on the full UA-Speech dataset and subsequently fine-tuned and evaluated on the SID sets of TORGO. This configuration assesses the model’s ability to adapt to a smaller corpus with distinct articulation patterns and recording environments.


\subsubsection{TORGO $\rightarrow$ UA-Speech}
\label{subsubsec:pretrained-torgo}
In the reverse configuration, DSSCNet is pre-trained on the TORGO dataset and then fine-tuned on the SID sets of UA-Speech. This setting presents a more challenging generalization task due to the larger speaker pool and greater variability in severity levels within UA-Speech.


This two-stage framework enables systematic evaluation of cross-corpus generalization and provides a strong initialization that enhances convergence and classification performance in both OSPS and LOSO settings.

\section{Results and Discussion}
\label{sec:results}
This section presents a detailed analysis of the performance of the proposed DSSCNet for SID dysarthric speech severity classification, evaluated under two distinct configurations: OSPS and LOSO. We examine and compare the model's performance across both settings and benchmark it against other methods. Furthermore, we conduct an ablation study to evaluate the contribution of individual architectural components to the overall performance. Also, we discuss the effectiveness of cross-corpus transfer learning, supported by t-SNE visualizations that illustrate the learned feature space and highlight DSSCNet’s ability to generalize across speakers and datasets.


\begin{figure*}[!ht]
    \centering
    \hfill
    \subfigure[\label{subfig:cnn-torgo-baseline}CNN + SE (Baseline) on TORGO]{\includegraphics[width=0.30\linewidth]{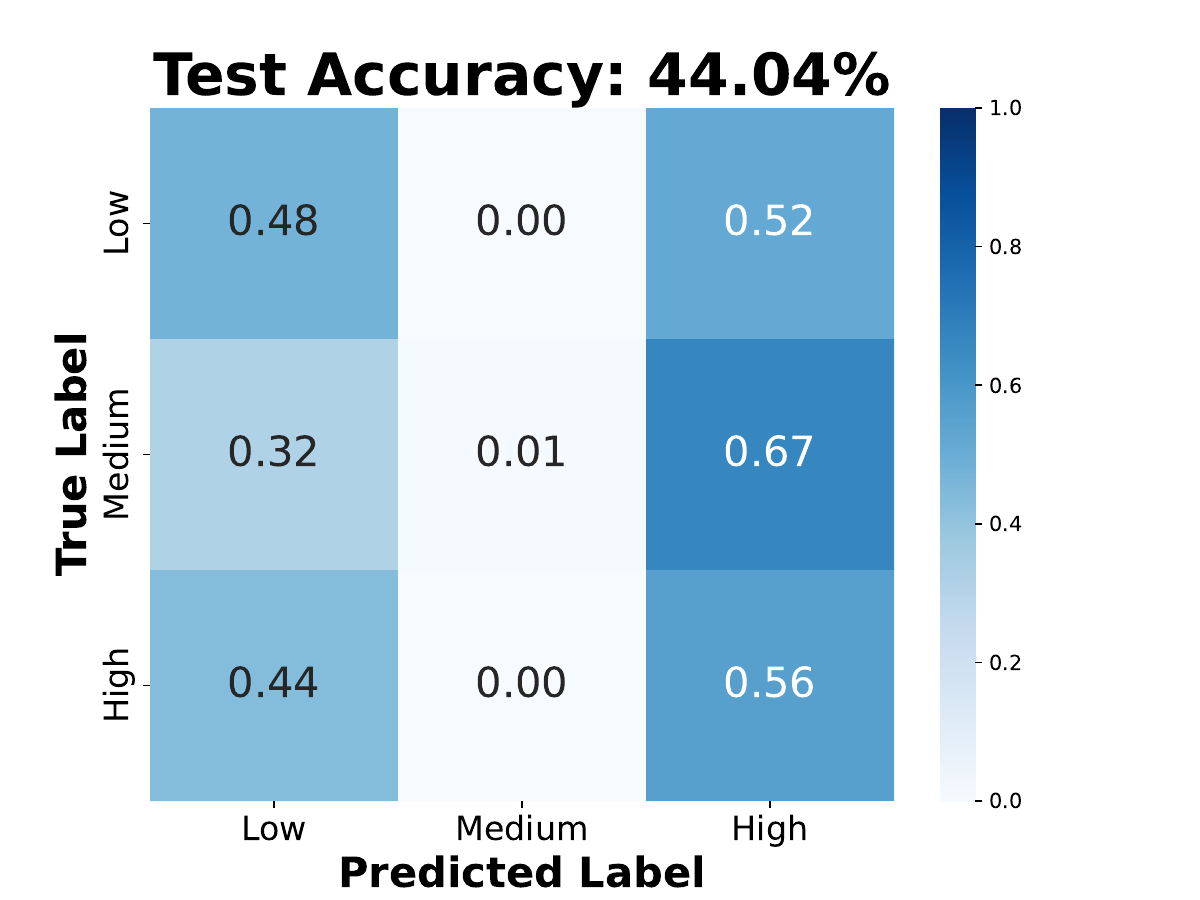}} \hfill
    \subfigure[\label{subfig:torgo-baseline}Proposed DSSCNet on TORGO]{\includegraphics[width=0.30\linewidth]{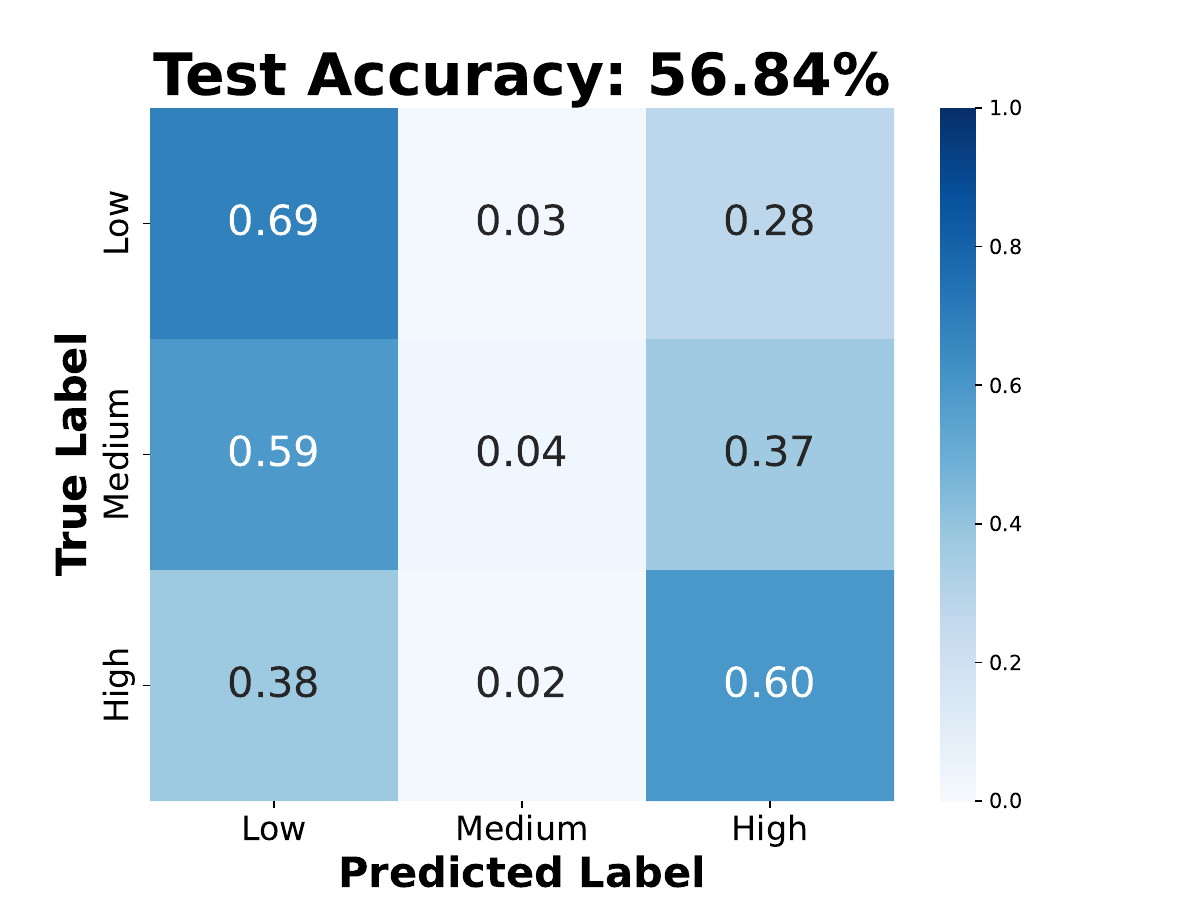}} \hfill
    \subfigure[\label{subfig:torgo-finetune}Fine-tuned DSSCNet on TORGO]{\includegraphics[width=0.30\linewidth]{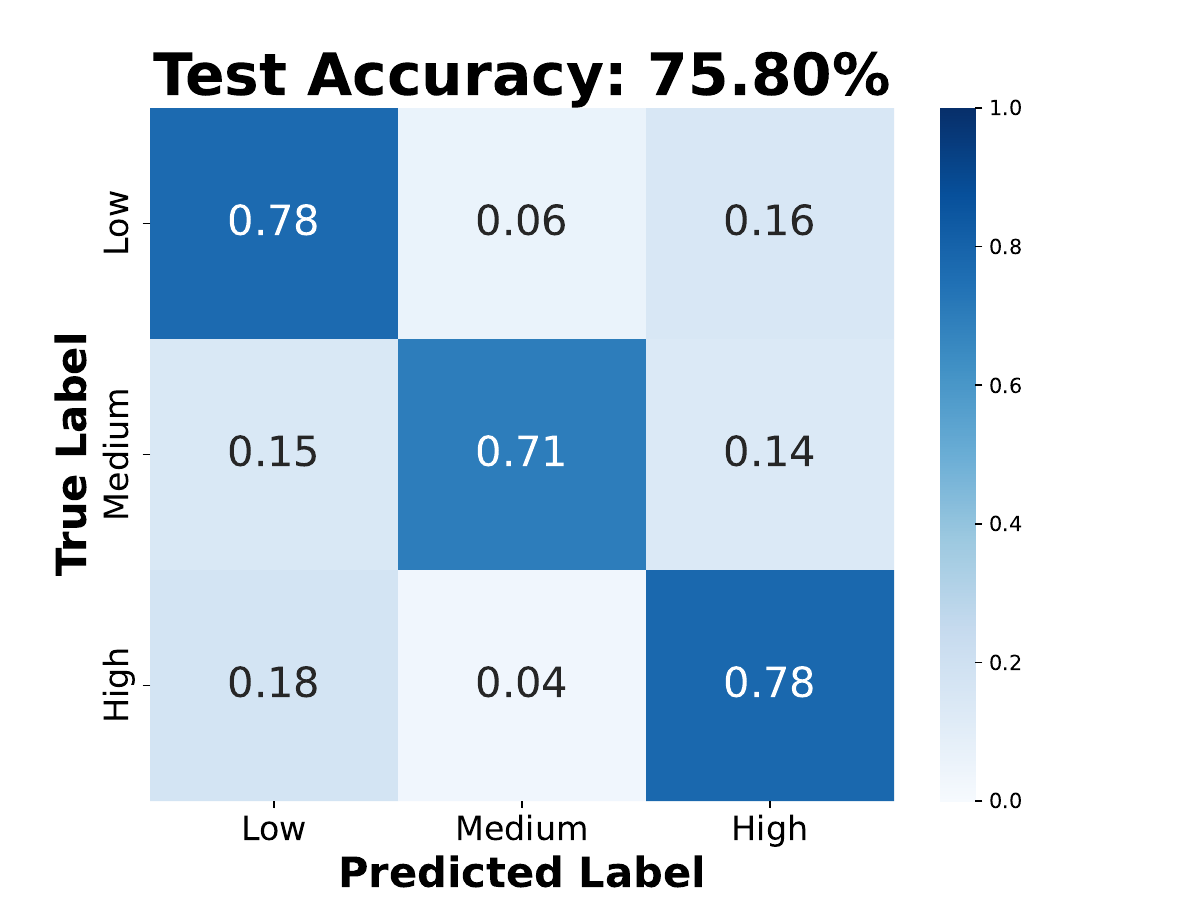}} \hfill
    \hfill

    \noindent\dotfill
    
    \hfill
    \subfigure[\label{subfig:cnn-uaspeech-baseline}CNN + SE (Baseline) on UA-Speech]{\includegraphics[width=0.30\linewidth]{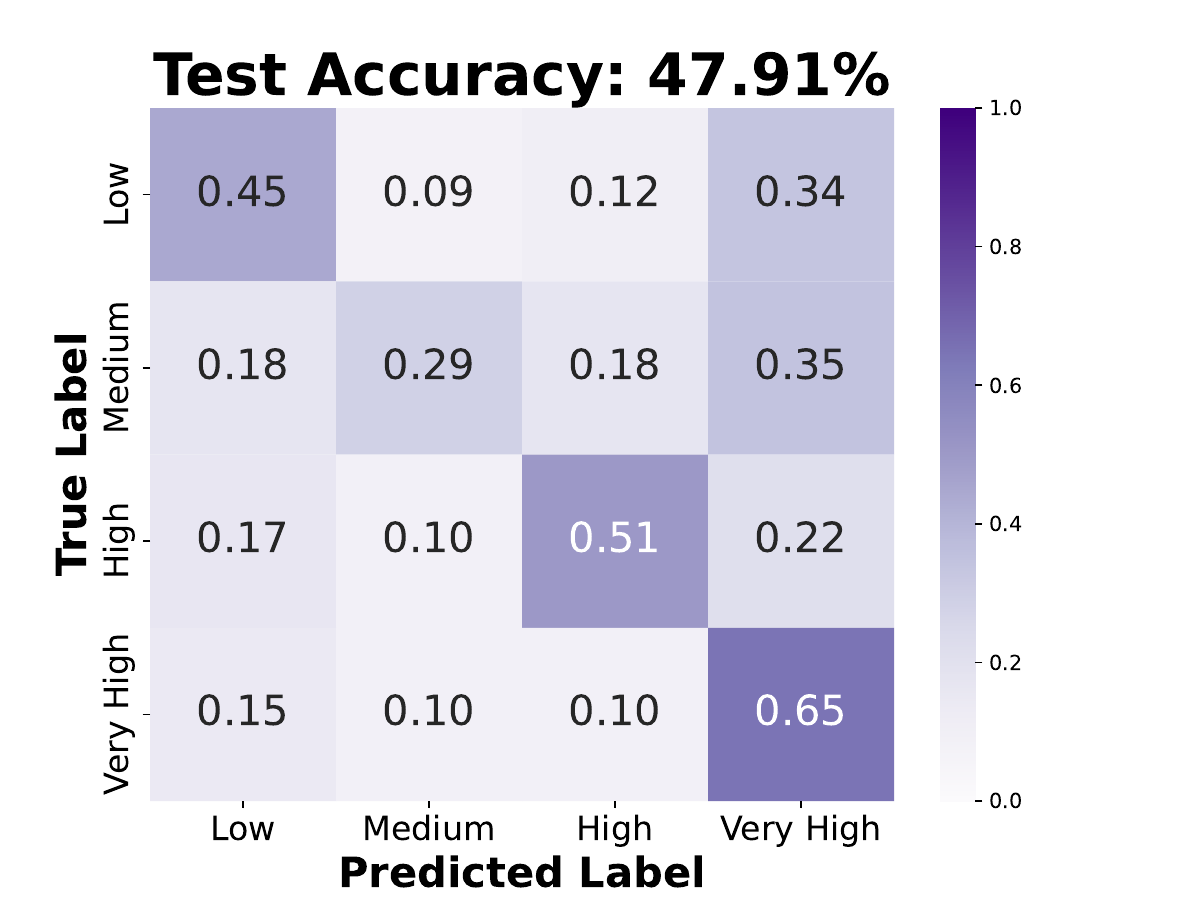}} \hfill
    \subfigure[\label{subfig:uaspeech-baseline}Proposed DSSCNet on UA-Speech]{\includegraphics[width=0.30\linewidth]{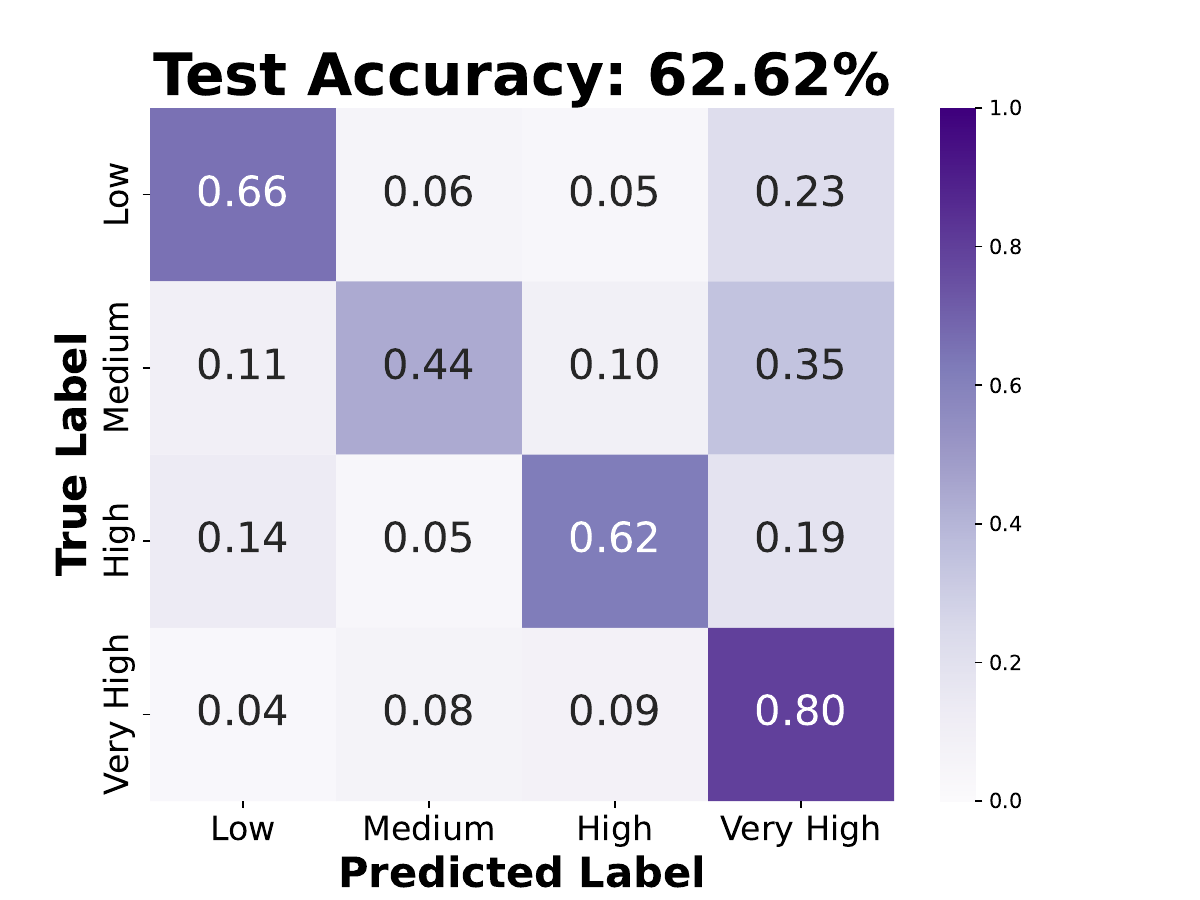}} \hfill
    \subfigure[\label{subfig:uaspeech-finetune}Fine-tuned DSSCNet on UA-Speech]{\includegraphics[width=0.30\linewidth]{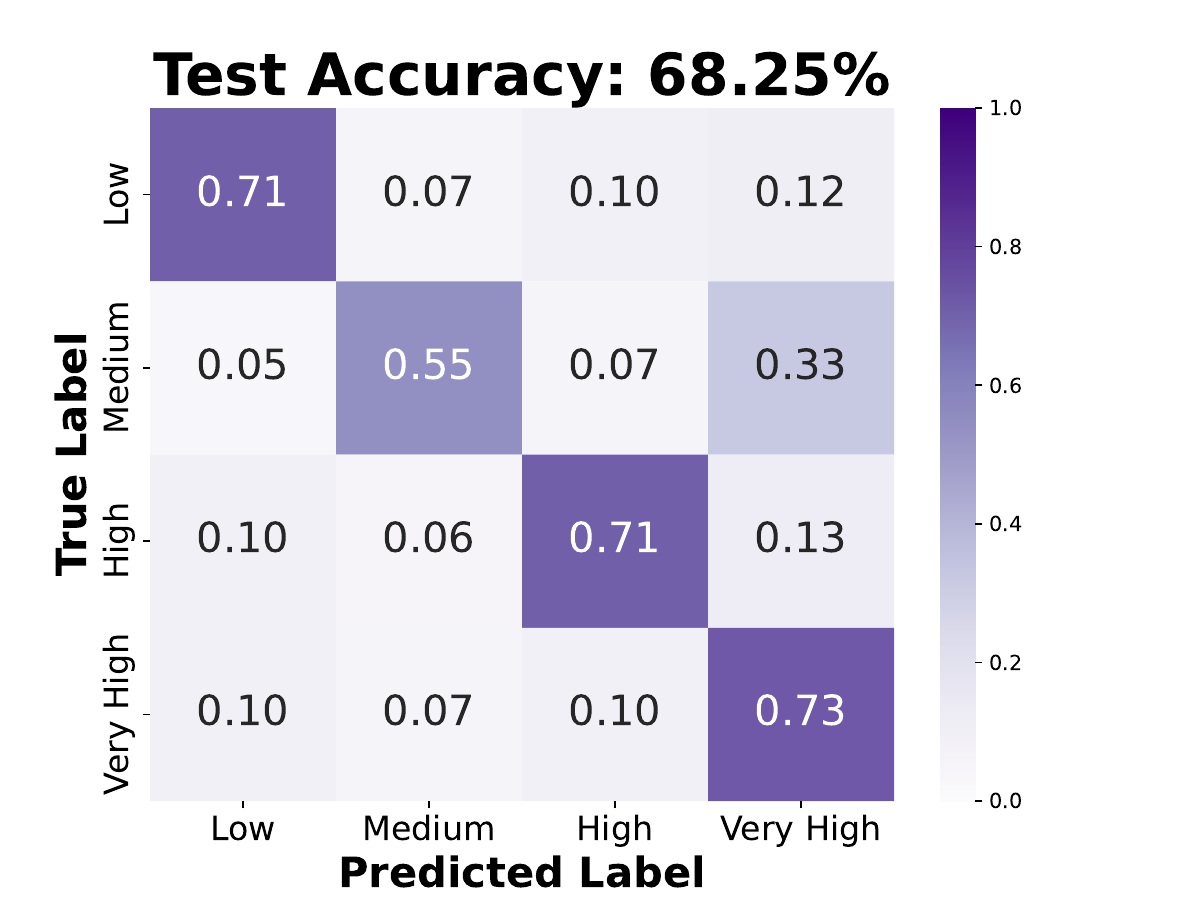}} \hfill
    \hfill
    \caption{Normalized confusion matrices illustrating multi-class dysarthric speech severity classification performance on the OSPS sets of TORGO using: (a) CNN + SE (Baseline), (b) DSSCNet, (c) DSSCNet with fine-tuning averaged across 18 sets, and of UA-Speech using: (d) CNN + SE (Baseline), (e) DSSCNet, (f) DSSCNet with fine-tuning averaged across 81 sets.}
    \label{fig:confusion-matrices}
\end{figure*}

\subsection{For One-Speaker-Per-Severity Setting}
\label{subsec:results-sid}

The generalization ability of DSSCNet in SID conditions is examined under the OSPS configuration using the TORGO and UA-Speech datasets. As shown in Fig. \ref{fig:confusion-matrices}, the normalized confusion matrices highlight the model’s capacity to maintain consistent severity classification across unseen speakers. DSSCNet demonstrates strong generalization in this setting, attributed to its architectural components such as residual connections and squeeze-excitation block which facilitate robust representation learning. Even prior to fine-tuning, the model achieves competitive performance, highlighting the strength of its architecture in capturing severity-related features. Fine-tuning further amplifies this capability, reinforcing DSSCNet’s effectiveness for dysarthric speech severity classification in real-world, SID scenarios.

On the \textbf{TORGO} dataset, the CNN + SE (baseline) model as discussed in Section \ref{subsubsec:loss-reduction} achieves an accuracy of 44.04\% (Fig. \ref{subfig:cnn-torgo-baseline}), demonstrating substantial difficulties in classifying dysarthric speech severities under unseen speakers. This relatively low performance can be attributed to the limited capacity of the baseline architecture, which lacks the depth and complexity required to capture nuanced acoustic patterns associated with varying severity levels. Furthermore, frequent misclassification of the Medium severity class is observed, largely due to class imbalance within the dataset as evident from Table \ref{tab:dataset}, which skews the model’s predictions toward more represented classes and reduces its sensitivity to Medium severity distinctions.

The proposed DSSCNet achieves an accuracy of 56.84\% (Fig. \ref{subfig:torgo-baseline}), an absolute improvement of 12.80\% over the baseline configuration. Key observations include:

\begin{itemize}
    \item \textbf{Low Severity:}
    \begin{itemize}
        \item 69\% of dysarthric speech samples labeled as Low severity were correctly classified.
        \item Only 3\% were misclassified as Medium severity.
        \item 28\% were misclassified as High severity, suggesting that Low and High severity speech patterns may share overlapping spectral characteristics in SID settings.
    \end{itemize}

    \item \textbf{Medium Severity:}
    \begin{itemize}
        \item Only 4\% of speech samples labeled as Medium severity were correctly classified.
	\item The majority of Medium severity samples were misclassified as Low (59\%) or High (37\%), indicating significant confusion.
	\item This poor performance is primarily attributed to the limited number of Medium severity samples in the dataset (Table \ref{tab:dataset}), which impedes the model’s ability to learn class-specific features.
    \end{itemize}

    \item \textbf{High Severity:}
    \begin{itemize}
        \item 60\% of speech samples labeled as High severity were correctly classified.
        \item Only 2\% were misclassified as Medium severity.
	\item 38\% of High severity samples were misclassified as Low, reflecting the inherent challenge of distinguishing severely impaired speech from milder forms in the absence of speaker-specific cues.
    \end{itemize}
\end{itemize}



DSSCNet demonstrates strong capacity to model dysarthric speech severity, leveraging its residual architecture and channel recalibration mechanisms to capture both fine-grained spectral features and high-level severity patterns enabling the model achieve state-of-the-art performance under SID settings as presented in Table \ref{tab:sota-comparison}. Despite this, performance under SID settings particularly in the presence of inter-speaker variability and class imbalance reveals persistent challenges in establishing clear decision boundaries, especially for underrepresented severity levels. To address these limitations, we adopt the cross-corpus transfer learning strategy outlined in Section \ref{subsec:framework}. Specifically, as detailed in Section \ref{subsubsec:pretrained-uaspeech}, a DSSCNet model trained on the UA-Speech dataset is fine-tuned on the TORGO dataset. By transferring knowledge from a larger and more diverse corpus, the model gains a stronger foundation for generalizing to the data-scarce target domain, resulting in improved discrimination across severity levels even under SID constraints.

With fine-tuning, DSSCNet achieves an accuracy of \textbf{75.80\%} on TORGO (Fig. \ref{subfig:torgo-finetune}), representing an absolute improvement of 18.96\% over our proposed network and 31.76\% over the baseline CNN + SE model. Notable improvements include:

\begin{itemize}
    \item \textbf{Low severity:}
    \begin{itemize}
        \item 78\% of speech samples labeled as Low severity were correctly classified.
        \item Only 6\% were misclassified as Medium severity.
        \item 16\% misclassified as High severity: A reduction from 28\%, indicating enhanced discrimination between Low and High severities.
    \end{itemize}
    \item \textbf{Medium severity:}
    \begin{itemize}
        \item Remarkable improvement in classifying the speech samples as Medium severity: improved from 4\% $\rightarrow$ 71\%.
        \item Misclassification reduces in both Low severity (59\% $\rightarrow$ 15\%) and High severity (37\% $\rightarrow$ 14\%).
        \item Improvement attributed to spectral feature transfer from pre-training, compensating for limited Medium severity samples (Table~\ref{tab:dataset}).
    \end{itemize}
    \item \textbf{High severity:}
    \begin{itemize}
        \item 78\% of speech samples labeled as High severity were correctly classified.
        \item Only 4\% were misclassified as Medium severity.
        \item Misclassification reduces in Low severity (38\% $\rightarrow$ 18\%).
    \end{itemize}
\end{itemize}

On the \textbf{UA-Speech} dataset, the CNN + SE (baseline) model achieves an accuracy of 47.91\% (Fig. \ref{subfig:cnn-uaspeech-baseline}), highlighting overall low performance under unseen speaker classification of dysarthric speech severity. The proposed DSSCNet achieves an accuracy of 62.62\% (Fig. \ref{subfig:uaspeech-baseline}) with an absolute improvement of 14.71\% over the baseline configuration. The findings are summarized as:

\begin{itemize}
    \item \textbf{Low severity:}
    \begin{itemize}
        \item 66\% of speech samples labeled as Low severity were correctly classified.
        \item Only 6\% and 5\% were misclassified as Medium and High severity respectively.
        \item 23\% were misclassified as Very High severity, suggesting spectral similarities between mild and severe dysarthric speech patterns.
    \end{itemize}

    \item \textbf{Medium severity:}
    \begin{itemize}
        \item 44\% of speech samples labeled as Medium severity were correctly classified, indicating considerable difficulty in differentiating Medium severity.
        \item 11\% and 10\% were misclassified as Low and High severity respectively.
        \item 35\% were misclassified as Very High severity.
    \end{itemize}

    \item \textbf{High severity:}
    \begin{itemize}
        \item 62\% of speech samples labeled as High severity were correctly classified.
        \item Only 5\% were misclassified as Medium severity.
        \item 14\% and 19\% were misclassified as Low and Very High severity respectively, demonstrating confusion between adjacent severity levels.
    \end{itemize}

    \item \textbf{Very High severity:}
    \begin{itemize}
        \item 80\% of speech samples labeled as Very High severity.
        \item Only 4\% were misclassified as Low severity.
        \item 8\% and 9\% were misclassified as Medium and High severity respectively.
    \end{itemize}
\end{itemize}

DSSCNet exhibits strong capability in modeling dysarthric speech severity, effectively utilizing its residual connections and squeeze-and-excitation mechanisms to extract discriminative features across varying severity levels aiding the model achieve state-of-the-art performance (Table \ref{tab:sota-comparison}). The model performs particularly well for the most distinct severity classes, capturing both subtle and pronounced articulatory deviations. However, under SID conditions, classification performance is affected by inter-speaker variability and the overlapping acoustic characteristics between adjacent severity levels.

To overcome the limitations observed in severity classification under SID settings, particularly for the Medium severity level, we apply the cross-corpus fine-tuning strategy described in Section \ref{subsec:framework}. In this configuration, the DSSCNet model trained on the TORGO dataset is fine-tuned and evaluated on the SID sets of UA-Speech as detailed in Section \ref{subsubsec:pretrained-torgo}. The pre-training phase allows the model to learn generalized dysarthric speech representations from a different corpus, which aids in capturing acoustic variability and improving class discrimination when transferred to the target domain.

Following fine-tuning, DSSCNet achieves a significantly improved accuracy of \textbf{68.25\%} on UA-Speech (Fig. \ref{subfig:uaspeech-finetune}), marking an 5.63\% absolute improvement over over DSSCNet without fine-tuning and an 20.34\% absolute improvement over the baseline. Detailed improvements observed in severity-level classification include:

\begin{itemize}
    \item \textbf{Low severity:}
    \begin{itemize}
        \item 71\% of speech samples labeled as Low severity were correctly classified.
        \item 7\% and 10\% were misclassified as Medium and High severity respectively.
        \item Misclassification reduces in Very High severity (23\% $\rightarrow$ 12\%), indicating improved severity differentiation.
    \end{itemize}

    \item \textbf{Medium severity:}
    \begin{itemize}
        \item 55\% of speech samples labeled as Medium severity were correctly classified.
        \item Only 5\% and 7\% were misclassified as Low and High severity respectively.
        \item Misclassification reduces in Very High severity (35\% $\rightarrow$ 33\%).
    \end{itemize}

    \item \textbf{High severity:}
    \begin{itemize}
        \item 71\% of speech samples labeled as High severity were correctly classified.
        \item Only 6\% were misclassified as Medium severity.
        \item 10\% and 13\% were misclassified as Low and Very High severity respectively.
    \end{itemize}

    \item \textbf{Very High severity:}
    \begin{itemize}
        \item 73\% of speech samples labeled as Very High severity were correctly classified.
        \item Only 7\% were misclassified as Medium severity.
        \item 10\% were misclassified in both Low and High severity respectively.
    \end{itemize}
\end{itemize}

These outcomes emphasize the effectiveness of cross-corpus fine-tuning in improving the generalizability of DSSCNet, particularly addressing challenges related to inter-speaker variability and severity-class imbalance.

\subsection{For Leave-One-Speaker-Out Setting}
\label{subsubsec:results-loso}
Table \ref{tab:results-loso} presents the average accuracy of the CNN + SE (baseline) model and the proposed DSSCNet under the LOSO evaluation protocol for both TORGO and UA-Speech datasets. The results are aggregated across 8 speakers for TORGO and 12 speakers for UA-Speech, each corresponding to a held-out speaker in the test set.

{
\renewcommand{\tabcolsep}{18pt}
\renewcommand{\arraystretch}{1.2}
\begin{table}[!ht]
    \centering
    \caption{Performance of DSSCNet on LOSO Setting of dysarthric speech severity classification on both TORGO and UA-Speech dataset. \cmark refers to fine-tuned performance.}
    \label{tab:results-loso}
    \begin{tabular}{c|c|c}
        \hline

        \hline
        \multirow{2}{*}{\textbf{Method}} & \multicolumn{2}{c}{\textbf{Accuracy (\%) on}} \\ \cline{2-3}
        & \textbf{TORGO} & \textbf{UA-Speech} \\
        \hline
        CNN + SE (baseline) & 54.66 & 56.84 \\
        Proposed DSSCNet & 63.47 & 64.18 \\
        Proposed DSSCNet \cmark & \textbf{77.76} & \textbf{79.44} \\
        \hline
        
        \hline
    \end{tabular}
\end{table}
}

The CNN + SE (baseline) achieves 54.66\% accuracy on TORGO and 56.84\% on UA-Speech. These results reflect the limited ability of the baseline model to generalize to unseen speakers, largely due to its shallow architecture and lack of mechanisms for capturing long-range dependencies or adapting to inter-speaker variability.

In contrast, the proposed DSSCNet demonstrates substantially improved performance, achieving 63.47\% on TORGO and 64.18\% on UA-Speech, an absolute improvement of 8.81\% and 7.34\%, respectively. This gain reflects DSSCNet’s enhanced capacity to extract informative features through its residual blocks, which preserve hierarchical representations, and its squeeze-and-excitation modules, which recalibrate channel-wise activations to focus on salient speech characteristics. The model shows strong generalization to unseen speakers, underscoring the effectiveness of its architectural design in handling inter-speaker variability and capturing severity-specific patterns in dysarthric speech.

The performance further improves significantly after cross-corpus fine-tuning, reaching \textbf{77.76\%} on TORGO and \textbf{79.44\%} on UA-Speech, an absolute improvement of 14.29\% and 15.26\% respectively. These gains underscore the effectiveness of leveraging learned dysarthric speech representations from a separate corpus to enhance model generalization.

Overall, the results confirm that while the baseline model provides a minimal benchmark, it lacks the architectural depth and transferability needed for robust severity classification in SID settings. The proposed DSSCNet, particularly when enhanced through cross-corpus fine-tuning, demonstrates substantial improvements and better suitability for real-world dysarthric speech applications.

\subsection{Ablation Study}
\label{subsec:ablation-study}
We conducted a component-wise ablation study under the OSPS setting on the TORGO dataset to examine the individual contributions of each architectural module in DSSCNet. Specifically, we evaluate the performance impact of removing or modifying the Convolutional Neural Network (CNN) layers, the Squeeze-and-Excitation (SE) block, and the Residual Blocks (RB). Each configuration (C2 to C6) represents a variant of the full model (C1) with one or more components altered. The results for these experiments are tabulated in Table \ref{tab:component-study}. This analysis helps determine which components are critical for effective dysarthric speech severity classification.

{
\renewcommand{\arraystretch}{1.6}
\begin{table}[!ht]
    \centering
    \caption{Component-wise study of DSSCNet using OSPS setting on TORGO dataset. Note that Config C1 denotes our best DSSCNet; CNN stands for Convolutional Neural Network, SE stands for Squeeze-and-Excitation Block, RB stands for Residual Block respectively.}
    \label{tab:component-study}
    \scalebox{0.7}{
    \begin{tabular}{c|ccc|c|c|c}
    \hline

    \hline
    \multicolumn{1}{l|}{\multirow{2}{*}{\textbf{Config}}} & \multicolumn{3}{c|}{\textbf{Components}}                                                                 & \multirow{2}{*}{\makecell{\textbf{Removed} \\ \textbf{Component(s)}}} & \multirow{2}{*}{\textbf{Accuracy (\%)}} & \multirow{2}{*}{\textbf{Loss}} \\ \cline{2-4}
    
    \multicolumn{1}{l|}{} & \multicolumn{1}{c|}{\textbf{CNN}} & \multicolumn{1}{c|}{\textbf{SE}} & \textbf{RB} & & & \\ \hline
    
    \textbf{C1} & \multicolumn{1}{c|}{\textbf{64, 128, 256}} & \multicolumn{1}{c|}{\textbf{256}} & \textbf{256, 512, 1024} & \textbf{-} & \textbf{56.84} & \textbf{0.82} \\ 
    
    C2 & \multicolumn{1}{c|}{64, 128, 256} & \multicolumn{1}{c|}{-} & 256, 512, 1024 & SE (256) & 51.99 (-4.85) & 1.20 (+0.38) \\ 
    
    C3 & \multicolumn{1}{c|}{128, 256} & \multicolumn{1}{c|}{256} & 256, 512, 1024 & CNN (64) & 49.63 (-7.21) & 1.09 (+0.27) \\ 
    
    C4 & \multicolumn{1}{c|}{256} & \multicolumn{1}{c|}{256} & 256, 512,  & CNN (64, 128) & 50.47 (-6.37) & 0.94 (+0.12) \\ 
    
    C5 & \multicolumn{1}{c|}{64, 128, 256} & \multicolumn{1}{c|}{256} & 512, 1024 & RB (256) & 43.26 (-13.58) & 1.03 (+0.23) \\ 
    
    C6 & \multicolumn{1}{c|}{64, 128, 256} & \multicolumn{1}{c|}{256} & 1024 & RB (256, 512) & 46.36 (-10.48) & 0.98 (+0.16) \\ 
    \hline

    \hline
\end{tabular}}
\end{table}
}

As shown in Table \ref{tab:component-study}, the full DSSCNet configuration (C1) achieves the highest accuracy of 56.84\% and the lowest cross-entropy loss of 0.82, confirming the effectiveness of its combined architecture. Removing the SE block (C2) leads to a significant performance drop to 51.99\%, highlighting the importance of channel-wise recalibration in focusing the network on salient speech features. A more drastic reduction in accuracy is observed when initial convolutional layers are removed (C3 and C4), demonstrating their role in capturing low-level spectral patterns essential for downstream classification. Similarly, configurations C5 and C6 show that the absence of either residual block stack leads to substantial declines in accuracy (43.26\% and 46.36\%, respectively), indicating that residual learning is crucial for preserving and transforming feature hierarchies.



\begin{figure}[!ht]
    \centering
    \includegraphics[width=\linewidth]{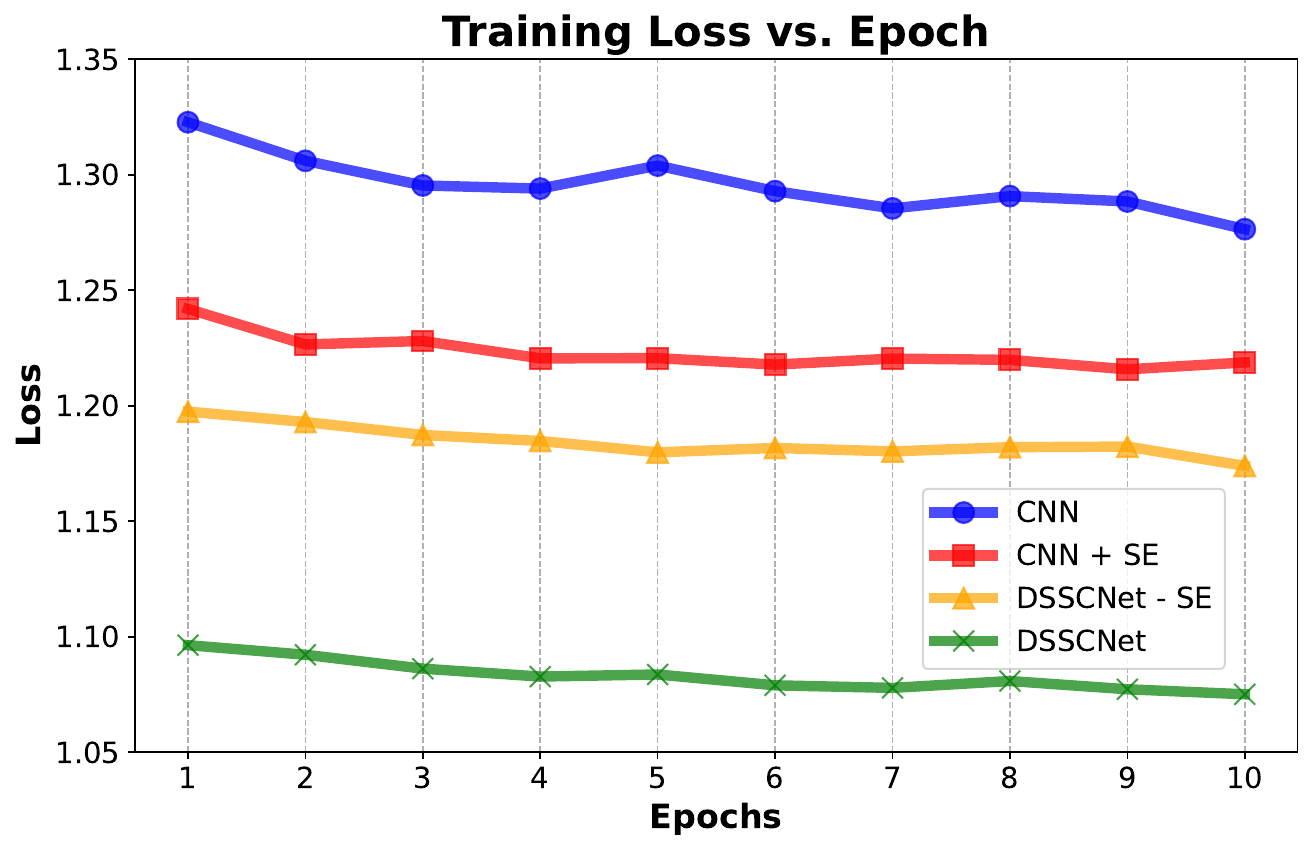}
    \caption{Training loss curves comparing CNN, CNN + SE, DSSCNet without SE (DSSCNet - SE), and the full DSSCNet configuration on the TORGO dataset.}
    \label{fig:loss-comparison}
\end{figure}

To further analyze the impact of SE block on model optimization, Fig. \ref{fig:loss-comparison} presents the loss curves across epochs for various configurations, including CNN, CNN + SE, DSSCNet without SE (DSSCNet - SE), and the full DSSCNet architecture. The plot clearly shows that architectures incorporating

the SE block: CNN + SE and full DSSCNet, exhibit faster and more stable convergence compared to their counterparts without SE. Notably, the full DSSCNet achieves the lowest loss throughout training, underscoring the synergy between residual learning and channel-wise recalibration. In contrast, the version of DSSCNet without SE converges more slowly and maintains a consistently higher loss, reinforcing the importance of the SE block in improving optimization efficiency and facilitating better feature discrimination for dysarthric speech severity classification.

We also conducted another experiment to assess the effectiveness of transfer learning as depicted in Fig. \ref{fig:framework} within our network. In this experiment, we compare DSSCNet against a baseline CNN + SE model on the OSPS setting of both TORGO and UA-Speech. The CNN + SE network, illustrated and discussed in Fig. \ref{fig:model-arch} and Section \ref{subsubsec:loss-reduction}, consists of a series of convolutional layers for hierarchical feature extraction, followed by SE blocks to refine channel-wise feature importance. The evaluation was performed both with and without fine-tuning, and the results are summarized in Table \ref{tab:accuracy-table}.

\renewcommand{\tabcolsep}{6pt}
\begin{table}[!ht]
    \centering
    \caption{Comparison of classification accuracies between DSSCNet and a simple CNN network with SE Block on SID sets of both TORGO (T) and UA-Speech (U) dataset. \cmark refers to fine-tuned performance.}
    \begin{tabular}{l|c|c|c|c}
        \hline
        
        \hline
        \multicolumn{1}{c|}{\textbf{Method}} & \textbf{Accuracy (\%)} & \textbf{Precision} & \textbf{Recall} & \textbf{F1-score}\\
        \hline
         CNN + SE (T) & 44.04 & 0.56 & 0.35 & 0.28 \\
         CNN + SE (T) \cmark & 52.37 & 0.55 & 0.41 & 0.35 \\

         DSSCNet (T) & 56.84 & 0.44 & 0.44 & 0.37  \\
         DSSCNet (T) \cmark & \textbf{75.80} & \textbf{0.76} & \textbf{0.75} & \textbf{0.76} \\
         \hline
         
         CNN + SE (U) & 47.91 & 0.48 & 0.47 & 0.46 \\
         CNN + SE (U) \cmark & 54.12 & 0.56 & 0.53 & 0.52 \\

         DSSCNet (U) & 62.62 & 0.66 & 0.62 & 0.63 \\
         DSSCNet (U) \cmark & \textbf{68.25} & \textbf{0.68} & \textbf{0.67} & \textbf{0.68} \\
         \hline

         \hline
    \end{tabular}
    \label{tab:accuracy-table}
\end{table}

The results presented in Table \ref{tab:accuracy-table} highlight the influence of architectural design and transfer learning on dysarthric speech severity classification, evaluated across accuracy, precision, recall, and F1-score. The baseline CNN + SE model, which incorporates channel-wise recalibration via SE blocks, achieves accuracies of 44.04\% on TORGO and 47.91\% on UA-Speech. However, its performance remains suboptimal in terms of recall (0.35 and 0.47) and F1-score (0.28 and 0.46), indicating poor sensitivity and weak class-level balance. These limitations reflect the inability of shallow convolutional models to fully capture the nuanced acoustic variability inherent in dysarthric speech.

Fine-tuning the CNN + SE model yields modest improvements across both datasets. On TORGO, accuracy rises to 52.37\% and F1-score to 0.35, while on UA-Speech, accuracy improves to 54.12\% with an F1-score of 0.52. Although these gains suggest that transfer learning aids generalization, the baseline architecture continues to struggle with SID variability.

\begin{figure*}[!ht]
    \centering
    \subfigure[\label{subfig:tsne-torgo-cnn}CNN + SE (Baseline)]{\includegraphics[width=0.30\linewidth]{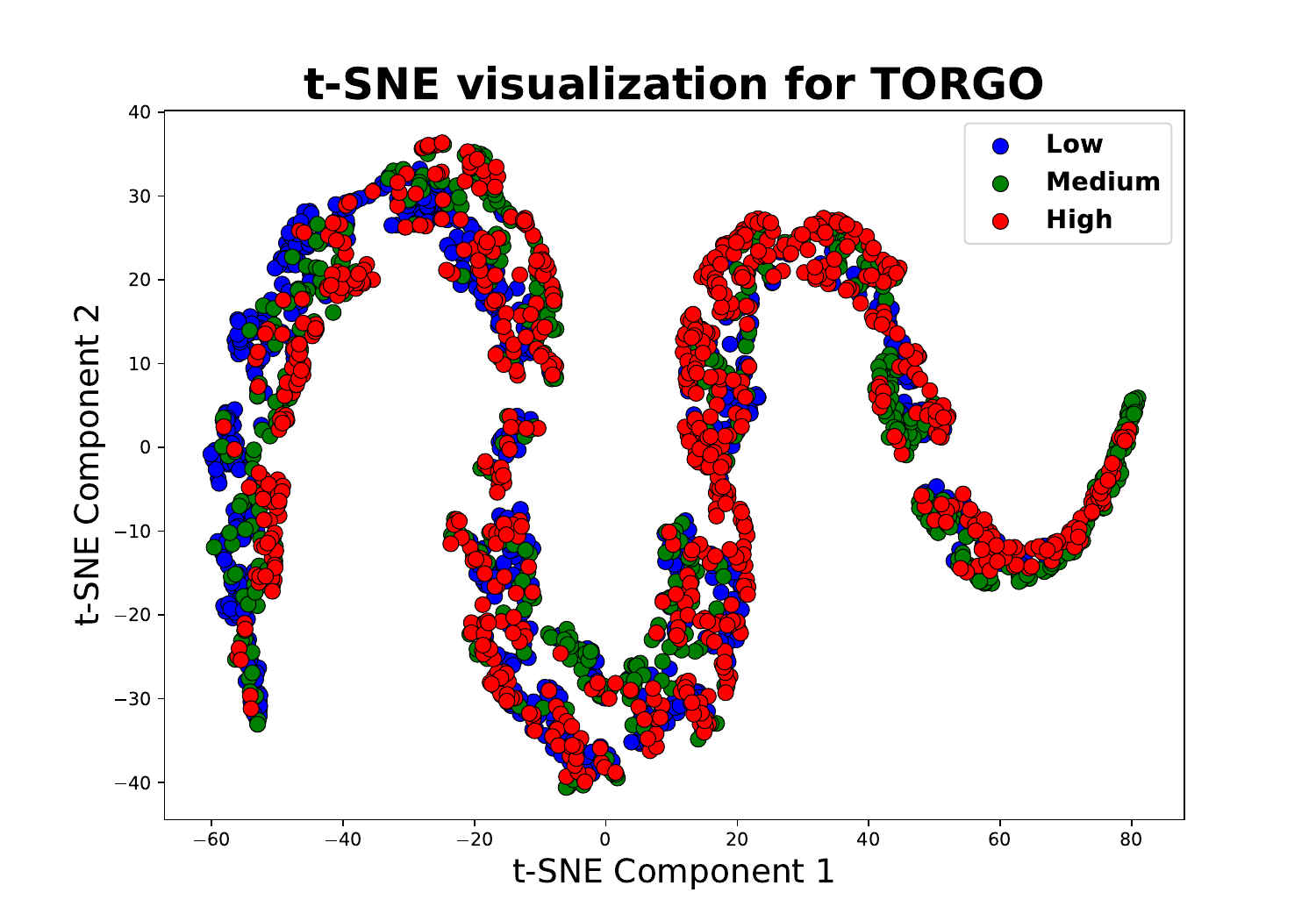}}
    \subfigure[\label{subfig:tsne-torgo-baseline}Proposed DSSCNet]{\includegraphics[width=0.30\linewidth]{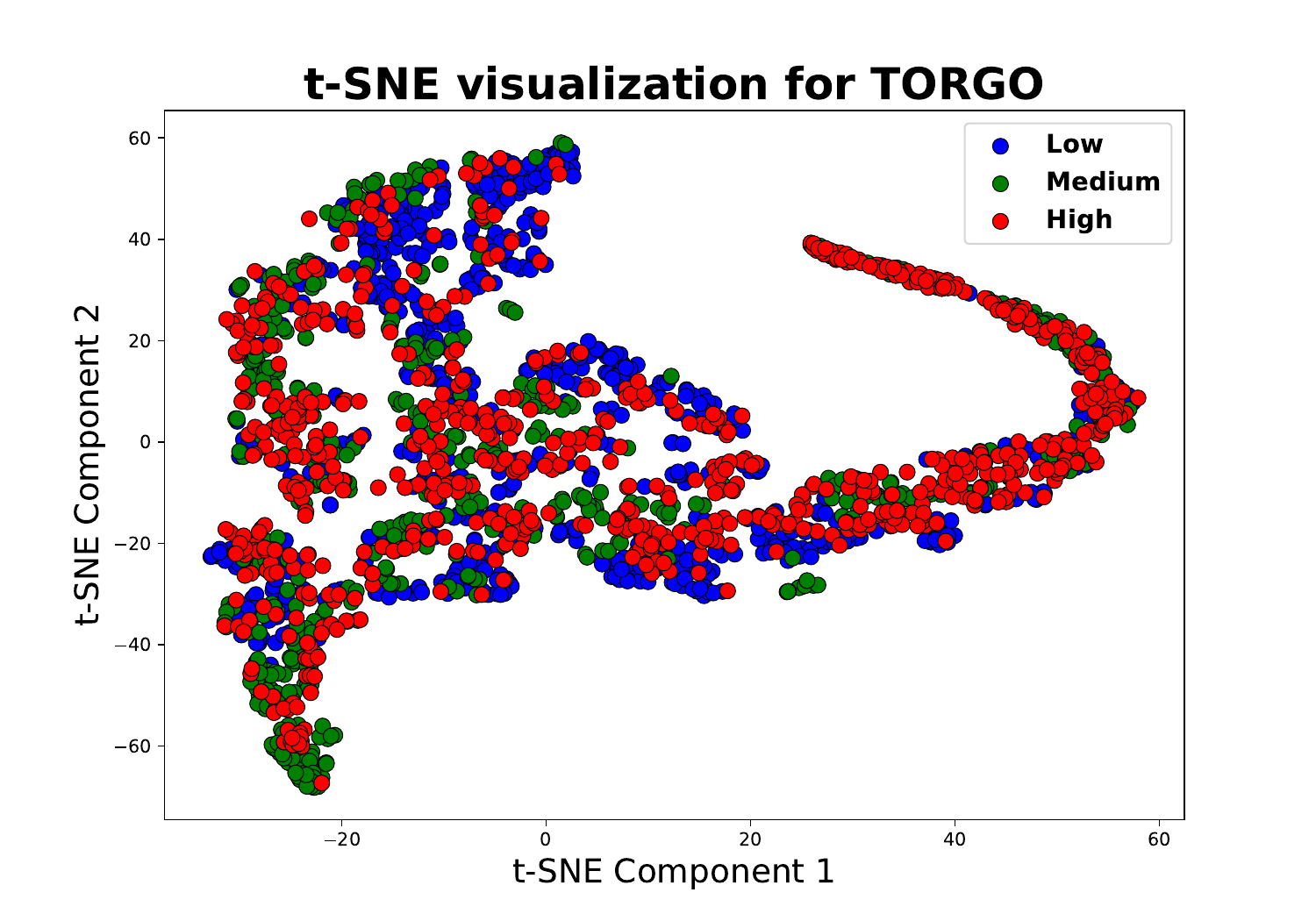}}
    \subfigure[\label{subfig:tsne-torgo-finetune}Fine-tuned DSSCNet]{\includegraphics[width=0.30\linewidth]{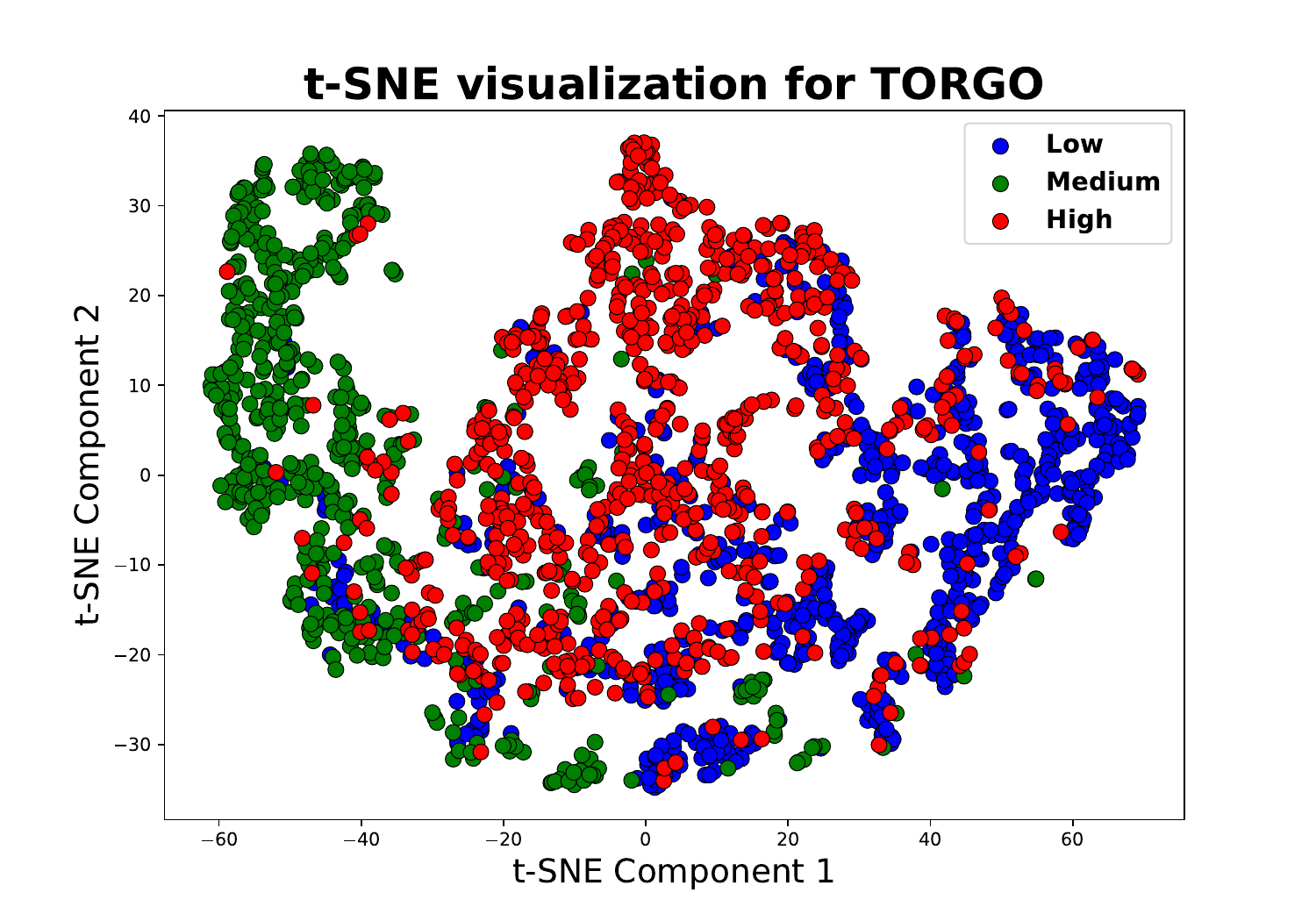}}
    \hfill
    
    \noindent\dotfill
    
    \subfigure[\label{subfig:tsne-uaspeech-cnn}CNN + SE (Baseline)]{\includegraphics[width=0.31\linewidth]{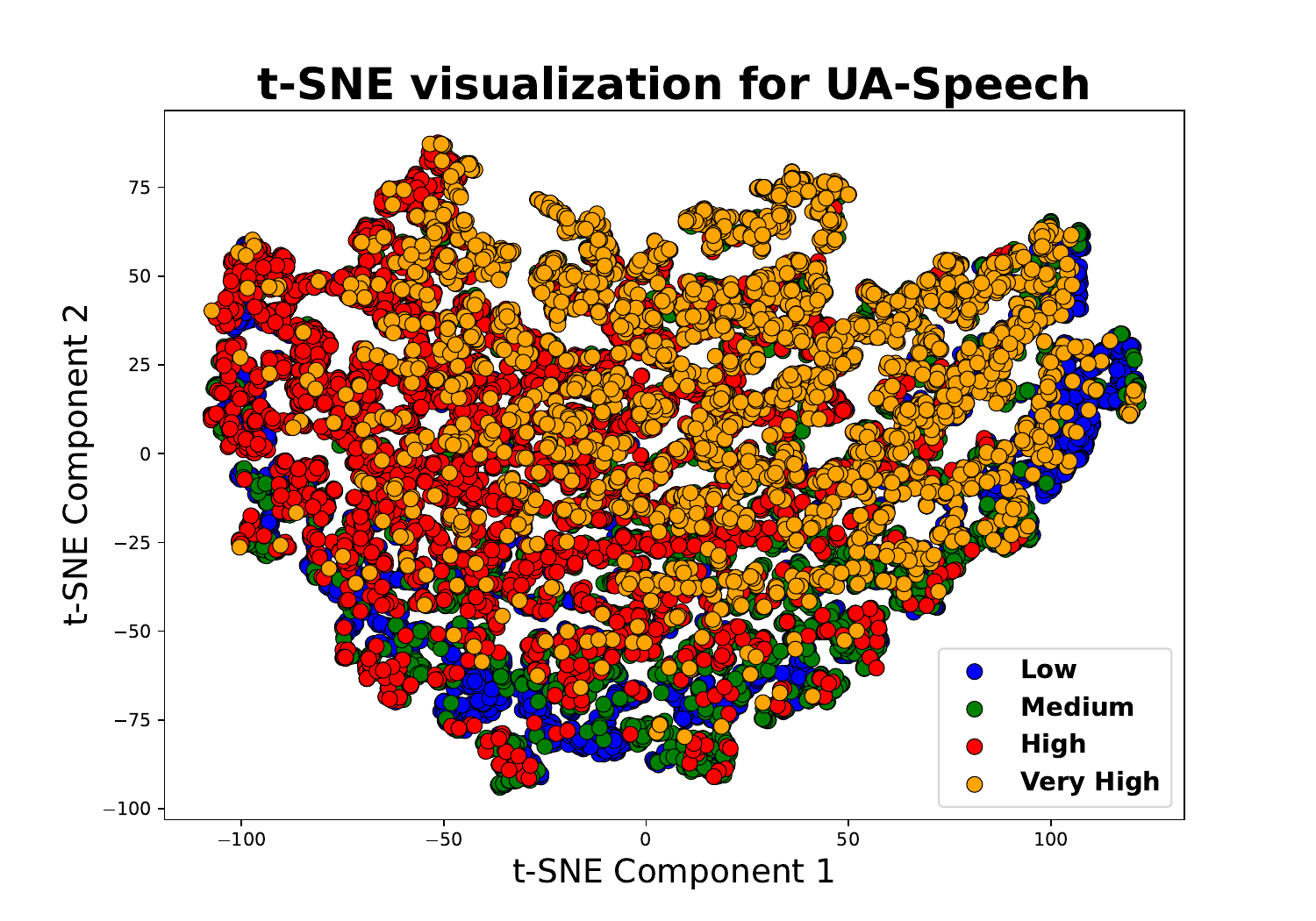}}
    \subfigure[\label{subfig:tsne-uaspeech-baseline}Proposed DSSCNet]{\includegraphics[width=0.31\linewidth]{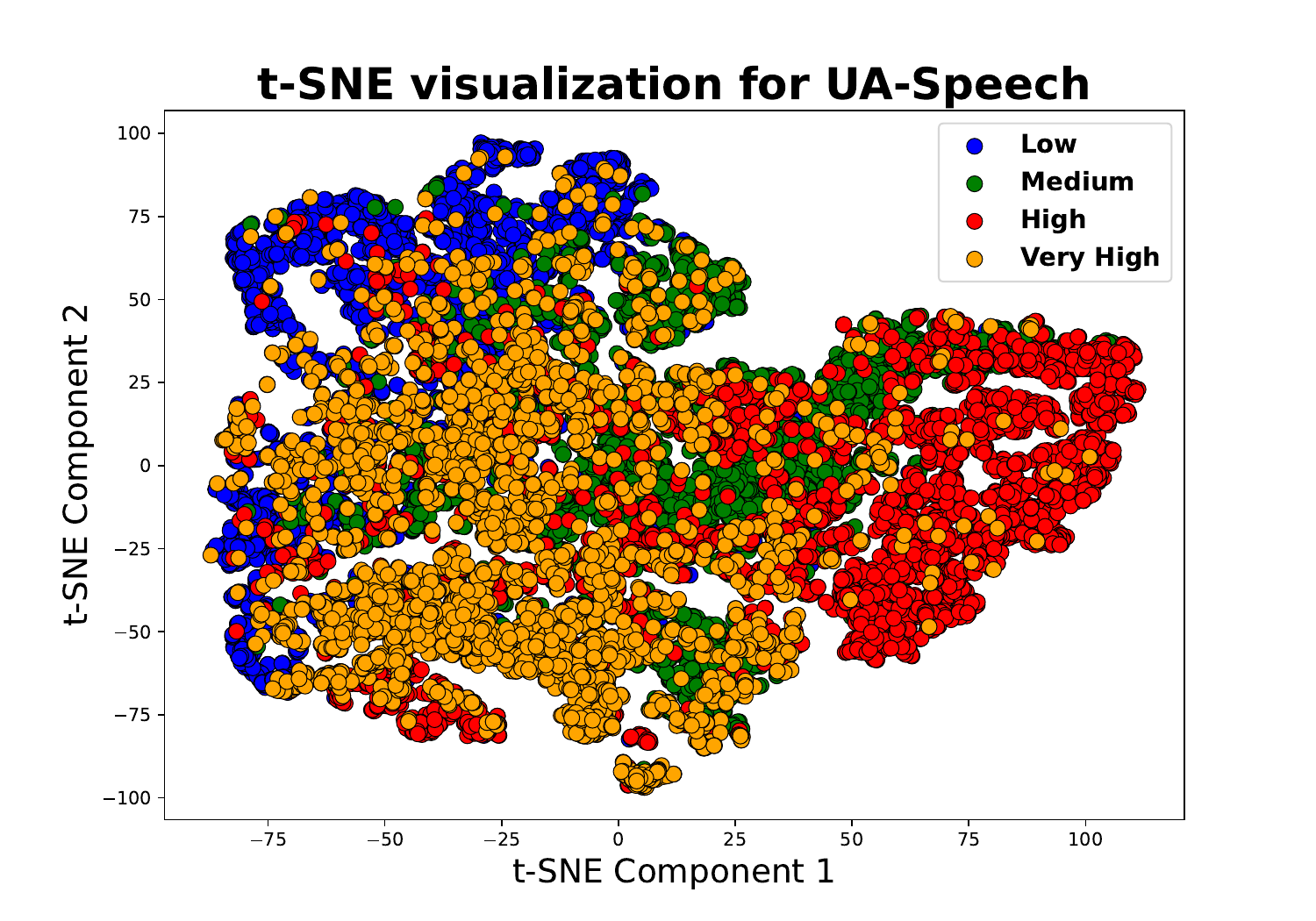}} 
    \subfigure[\label{subfig:tsne-uaspeech-finetune}Fine-tuned DSSCNet]{\includegraphics[width=0.31\linewidth]{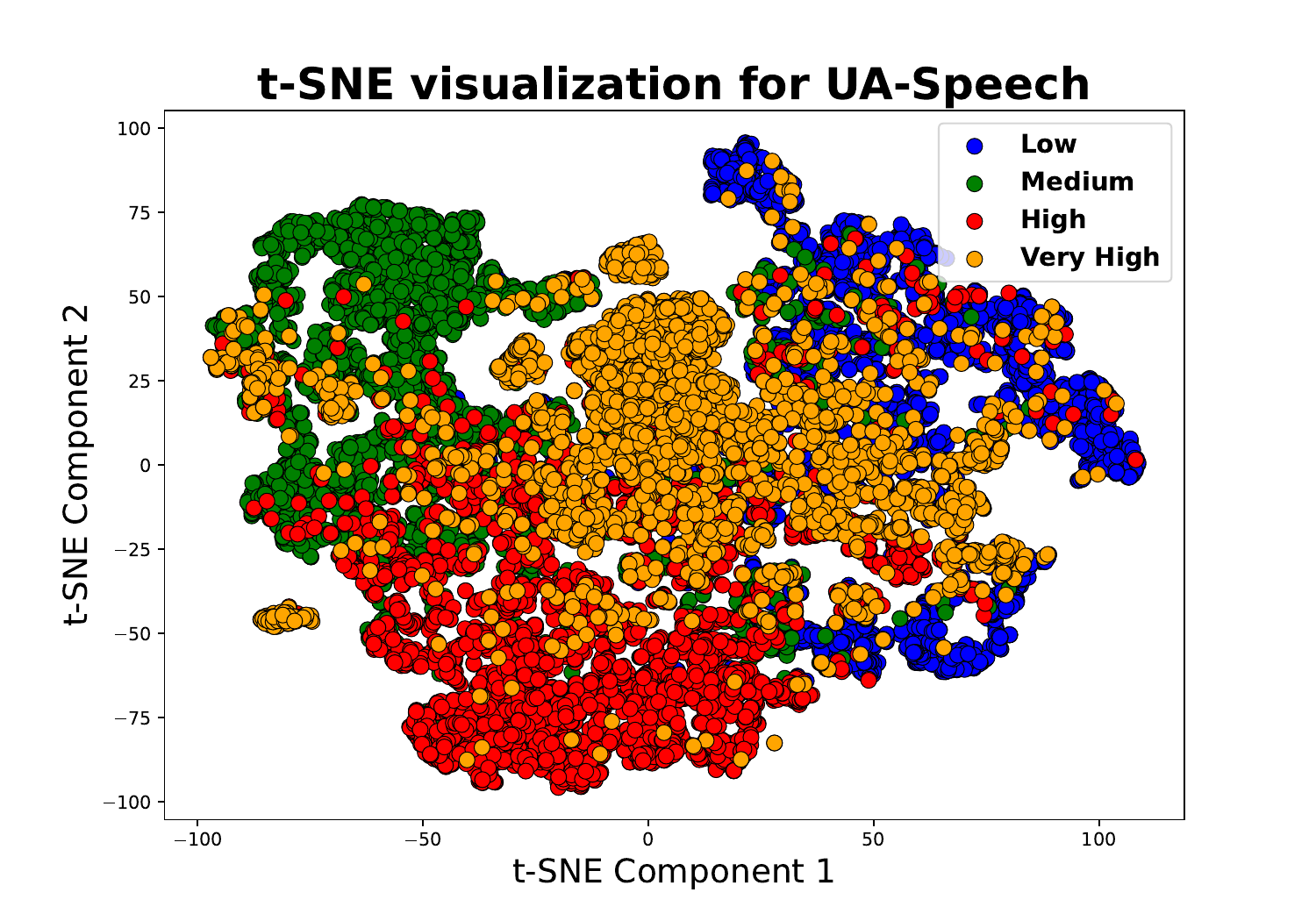}} 
    \caption{t-SNE plots of learned feature embeddings for dysarthria severity classification on TORGO using: (a) CNN + SE (Baseline), (b) DSSCNet, (c) DSSCNet with fine-tuning, and on UA-Speech using: (d) CNN + SE (Baseline), (e) DSSCNet, (f) DSSCNet with fine-tuning.}
    \label{fig:tsne}
\end{figure*}

In contrast, DSSCNet consistently outperforms the baseline across all metrics and datasets. The proposed network achieves 56.84\% accuracy on TORGO and 62.62\% on UA-Speech, with corresponding F1-scores of 0.37 and 0.63, highlighting the benefits of residual learning and deeper hierarchical structures in modeling severity-specific patterns. Fine-tuning further amplifies this performance: DSSCNet reaches 75.80\% accuracy and 0.76 F1-score on TORGO, and 68.25\% accuracy with 0.68 F1-score on UA-Speech. These results underscore the value of combining architectural depth with cross-corpus transfer learning for robust and generalizable severity classification in dysarthric speech.

These results collectively underscore the complementary benefits of deep residual architectures and cross-corpus transfer learning. While squeeze-and-excitation mechanisms contribute to improved feature weighting, it is the synergy between architectural depth and knowledge transfer that enables DSSCNet to achieve state-of-the-art performance across diverse evaluation metrics. The consistent improvements across accuracy, precision, recall, and F1-score reinforce the robustness and generalizability of the proposed approach in SID dysarthric speech severity classification.

\subsection{Effects of Cross-Corpus Learning}
\label{subsec:effects-cross-corpus}
Fig. \ref{fig:tsne} presents the t-SNE visualizations of learned feature embeddings from the penultimate layer of the respective models for both the TORGO and UA-Speech datasets. These plots provide an interpretable representation of how the models internally represent dysarthric speech severity levels under various architectural configurations and training strategies. Notably, the spatial separation among data points of different severity classes indicates the model’s capacity to learn discriminative embeddings, while the extent of overlap corresponds to confusion across severity levels.

For the TORGO dataset, Fig. \ref{subfig:tsne-torgo-cnn} shows that the CNN + SE (baseline) model yields poorly separated clusters, with significant overlap among the severity classes, indicating limited capacity for feature discrimination. In Fig. \ref{subfig:tsne-torgo-baseline}, DSSCNet without fine-tuning demonstrates improved class-wise clustering, although considerable overlap remains, especially between Medium and High severity classes. After applying cross-corpus fine-tuning, Fig. \ref{subfig:tsne-torgo-finetune} shows that DSSCNet achieves markedly improved separability, with clearer boundaries between severity classes and denser within-class grouping. This enhanced embedding structure aligns with the improved classification performance observed in the confusion matrices (Fig. \ref{subfig:torgo-baseline} and \ref{subfig:torgo-finetune}), where misclassifications particularly for the Medium class are significantly reduced after fine-tuning.

A similar trend is observed on the UA-Speech dataset. The CNN + SE model (Fig. \ref{subfig:tsne-uaspeech-cnn}) exhibits widespread inter class overlap among the different severity levels. DSSCNet without fine-tuning (Fig. \ref{subfig:tsne-uaspeech-baseline}) begins to separate classes more clearly, although Medium severity samples remain dispersed and intermingled with adjacent levels. Fine-tuned DSSCNet embeddings shown in Fig. \ref{subfig:tsne-uaspeech-finetune}, demonstrate substantially more compact and well separated clusters, especially for the Low, High and Very High classes. This qualitative improvement in feature space organization is corroborated by the confusion matrices in Fig. \ref{subfig:uaspeech-baseline} and \ref{subfig:uaspeech-finetune}, where fine-tuning improves classification accuracy and reduces cross-class confusion, particularly in the Medium and High categories.

These visualizations confirm that cross-corpus learning not only enhances classification metrics but also leads to more structured and semantically meaningful internal representations, which are crucial for speaker-independent generalization in dysarthric speech severity classification.

\subsection{Comparison with Previous Works}
\label{subsec:comparison}

\renewcommand{\tabcolsep}{14pt}
\renewcommand{\arraystretch}{1.5}
\begin{table*}[!ht]
    \centering
    \caption{Speaker-Independent Classification Accuracy Comparison of the proposed DSSCNet with state-of-the-art methods on TORGO and UA-Speech.}
    \resizebox{\linewidth}{!}{
    \begin{tabular}{lllll}
        \hline
        
        \hline
           \multicolumn{1}{c}{\textbf{Methods}} & \multicolumn{1}{c}{\textbf{Feature}} & \multicolumn{1}{c}{\textbf{Data usage}} & \multicolumn{1}{c}{\textbf{Results}} & \multicolumn{1}{c}{\textbf{Remarks}} \\
        \hline
        \multirow{4}{*}{CNN \cite{8884185}} & \multirow{4}{*}{Mel-Spectrogram} & TORGO: 1358 words used for training and & & - For UA-Speech and TORGO, all the samples \\
        & & 339 used for testing & TORGO: 54.00\% (OSPS) & were not used \\
        & & UA-Speech: Only 455 words of each speaker & UA-Speech: 54.10\% (OSPS) & - LOSO cross-validation was not used for \\
        & & used with LOSO cross-validation & & TORGO \\
        
        \hline

        \multirow{4}{*}{CNN \cite{javanmardi2024pre}} & \multirow{4}{*}{HuBERT features} & TORGO: All the samples from all 8 & & - TORGO: the accuracy represents the average \\
        & & dysarthric speaker used & TORGO: 49.83\% (OSPS) & accuracy of 18 rounds \\
        & & UA-Speech: all the samples from only 12 & UA-Speech: 48.01\% (OSPS) & - UA-Speech: the accuracy represents the average \\
        & & dysarthric speakers used & & accuracy of 81 rounds \\
        
        \hline

        \multirow{3}{*}{DNN \cite{joshy2022automated}} & \multirow{3}{*}{MFCC-based} & Common words used for training and & \multirow{3}{*}{UA-Speech: 49.22\% (OSPS)} & \multirow{3}{*}{- Study utilized only the UA-Speech corpus } \\
        & & uncommon words used for testing & & \\
        & i-vectors & with LOSO cross-validation & & \\
        
        \hline

        \multirow{4}{*}{SVM \cite{9054492}} & & Common words used for training and & UA-Speech: & \multirow{4}{*}{- Study utilized only the UA-Speech corpus}\\
        & Deep Speech & uncommon words for testing for SD & 53.90\% (OSPS), & \\
        & posteriors & round-robin LOSO cross-validation & 65.20\% (LOSO), & \\
        & & for SID case & & \\

        \hline

        \multirow{4}{*}{SECNN \cite{joshy2023biomedical}} & \multirow{4}{*}{Mel-Spectrogram} & Training: 455 common words of 8 & & \multirow{4}{*}{- Study utilized only the UA-Speech corpus} \\
        & & speakers & UA-Speech: 57.23\% (OSPS) & \\
        & & Testing: uncommon words of the left & and 67.25\% (LOSO) & \\
        & & out speakers & & \\

        \hline \hline

        \multirow{4}{*}{\textbf{DSSCNet (Proposed)}} & \multirow{4}{*}{Mel-Spectrogram} & TORGO: All the samples from all 8 & TORGO: \textbf{56.84\%} (OSPS), & - TORGO: the accuracy represents the average \\
        & & dysarthric speaker used & and \textbf{63.47\%} (LOSO) & accuracy of 18 rounds \\
        & & UA-Speech: all the samples from 12 & UA-Speech: \textbf{62.62\%} (OSPS), & - UA-Speech: the accuracy represents the average \\
        & & dysarthric speakers used & and \textbf{64.18\%} (LOSO) & accuracy of 81 rounds \\ 
        
        \hline
        
        \multirow{4}{*}{\textbf{DSSCNet (Proposed)}} & \multirow{4}{*}{Mel-Spectrogram} & TORGO: All the samples from all 8 & TORGO: \textbf{75.80\%} (OSPS), & - TORGO: the accuracy represents the average \\
        & & dysarthric speaker used & and \textbf{77.76\%} (LOSO) & accuracy of 18 rounds \\
        \textbf{+ fine-tuning} & & UA-Speech: all the samples from 12 & UA-Speech: \textbf{68.25\%} (OSPS), & - UA-Speech: the accuracy represents the average \\
        & & dysarthric speakers used & and \textbf{79.44\%} (LOSO) & accuracy of 81 rounds \\

        \hline
        
        \hline
    \end{tabular}}
    \label{tab:sota-comparison}
\end{table*}

The performance of the proposed DSSCNet is evaluated against state-of-the-art architectures to assess its effectiveness in dysarthric speech severity classification. The comparison highlights DSSCNet’s ability to capture severity-related speech patterns, leveraging transfer learning and cross-corpus adaptation for improved SID performance.

Table \ref{tab:sota-comparison} compares DSSCNet with existing approaches under SID settings. Even without fine-tuning, DSSCNet demonstrates strong generalization capabilities under SID settings when compared to existing methods. On the TORGO dataset, it achieves an accuracy of 56.84\%, outperforming CNN (Mel-Spectrogram) \cite{8884185} at 54.00\% and CNN with HuBERT \cite{javanmardi2024pre} at 49.83\%, with absolute improvements of 2.84\% and 7.01\%, respectively. On UA-Speech, DSSCNet attains 62.62\%, marking a gain of 8.72\% over CNN with DeepSpeech \cite{9054492} at 53.90\% and 13.40\% over CNN with HuBERT \cite{javanmardi2024pre}. These findings underscore the effectiveness of DSSCNet’s architectural design in capturing severity-specific speech characteristics and generalizing across unseen speakers.

Upon fine-tuning, DSSCNet achieves an accuracy of 75.80\% on the TORGO dataset, representing an absolute improvement of 21.80\% over CNN (Mel-Spectrogram) \cite{8884185}, which attains 54.00\%, and a 25.97\% gain over CNN with HuBERT \cite{javanmardi2024pre}, which achieves 49.83\%. On the UA-Speech dataset, DSSCNet reaches 68.25\%, surpassing CNN with DeepSpeech \cite{9054492} by 14.35\% (from 53.90\%) and outperforming CNN with HuBERT by 20.24\%. It also exceeds the performance of a DNN model utilizing MFCC-based i-vectors \cite{joshy2022automated}, which records an accuracy of 49.22\%, by 19.03\%. These results highlight the effectiveness of the proposed fine-tuning strategy in enhancing model generalization and improving severity classification across different dysarthric speech datasets.

\section{Conclusion}
\label{sec:conclusion}

In this work, we proposed DSSCNet, a deep learning framework for dysarthric speech severity classification under SID scenarios. The model incorporates residual connections and squeeze-and-excitation (SE) blocks to enhance spectral feature discrimination and representation learning. DSSCNet was evaluated under two SID configurations: OSPS and LOSO across two benchmark datasets: TORGO and UA-Speech. To further improve generalization, we employed a cross-corpus transfer learning strategy, enabling the model to effectively adapt to unseen speakers and dataset-specific characteristics. Extensive experiments demonstrate that DSSCNet consistently outperforms baseline and existing state-of-the-art models across multiple performance metrics. The improvements are particularly significant for intermediate severity levels, which are commonly misclassified due to spectral overlap. These findings, reinforced through confusion matrix analysis and t-SNE visualizations, highlight the model’s robustness, generalization capability, and architectural efficiency for real-world dysarthric speech applications.
 
This study offers practical value in clinical assessment and assistive speech technologies. Accurate severity classification supports SLPs in developing personalized therapy plans and monitoring patient progress. Furthermore, the DSSCNet, a severity aware classification model, can enhance user adaptive ASR, ASV systems for users with dysarthria by dynamically adjusting model behavior based on severity. The cross-corpus transfer learning approach employed here promotes model scalability to new clinical datasets, enhancing robustness in real-world deployments where speaker characteristics and recording conditions are highly variable. Consequently, this work contributes towards the development of accessible, data-driven solutions that support individuals with speech impairments.

\bibliographystyle{IEEEtran}

\end{document}